\documentclass[alpha-refs]{wiley-article}
    
\usepackage{graphicx}
\graphicspath{{Figures/}}
\usepackage{caption}
\usepackage{subfig} 
\usepackage[space]{grffile}
\usepackage{latexsym}
\usepackage{textcomp}
\usepackage{longtable}
\usepackage{tabulary}
\usepackage{booktabs,array,multirow}
\usepackage{amsfonts,amsmath,amssymb}
\usepackage{natbib}
\usepackage{url}
\usepackage{hyperref}
\hypersetup{colorlinks=false,pdfborder={0 0 0}}
\usepackage{etoolbox}

\usepackage{lipsum}

\usepackage{siunitx}
\usepackage{hhline}
\usepackage{colortbl}
\usepackage{rotating}
\usepackage[linesnumbered,ruled,vlined]{algorithm2e}
\usepackage{array}
\newcolumntype{P}[1]{>{\centering\arraybackslash}p{#1}}
\usepackage{multicol}
\usepackage{threeparttable}
\usepackage{todonotes}

\newif\iflatexml\latexmlfalse

\AtBeginDocument{\DeclareGraphicsExtensions{.pdf,.PDF,.eps,.EPS,.png,.PNG,.tif,.TIF,.jpg,.JPG,.jpeg,.JPEG}}

\usepackage[utf8]{inputenc}
\usepackage[T1]{fontenc}

\usepackage[english]{babel}

\usepackage{siunitx}

\iflatexml

\else

\corraddress{Koorush Ziarati, Department of Engineering and Computer Science and Information Technology, Shiraz University, Shiraz, Iran}
\corremail{ziarati@shirazu.ac.ir}

\fundinginfo{Test} 
\fi

\papertype{Original Article}

\title{Graph-based Local Climate Classification in Iran}

\author[1]{Neda Akrami}
\author[2]{Koorush Ziarati}
\author[3]{Soumyabrata Dev}

\affil[1]{Department of Engineering and Computer Science and Information Technology, Shiraz University, Shiraz, Iran}
\affil[2]{Department of Engineering and Computer Science and Information Technology, Shiraz University, Shiraz, Iran}
\affil[3]{ADAPT SFI Research Centre, University College Dublin, Ireland}

\runningauthor{Akrami \textit{et al.}}

\makeatletter 
\patchcmd{\wiley@affilmetadata}%
  {\textbf{Funding information}\\
   \@fundinginfo\par
  }%
  {\par}%
  {}
  {}
\makeatother 

\begin{document}

\maketitle

\begin{abstract}
In this paper, we introduce a novel graph-based method 
to classify the regions with similar climate in a local area. We refer our proposed method as Graph Partition Based Method (GPBM). Our proposed method attempts to overcome the shortcomings of the current state-of-the-art methods in the literature. 
It has no limit on the number of variables that can be used and also preserves the nature of climate data. To illustrate the capability of our proposed algorithm, we benchmark its performance with other state-of-the-art climate classification techniques. 
The climate data is collected from $24$ synoptic stations in Fars province in southern Iran. The data includes seven climate variables stored as time series from 1951 to 2017. Our results exhibit that our proposed method performs a more realistic climate classification with less 
computational time. It can save more information during the climate classification process and is therefore efficient in further data analysis. Furthermore, using our method, we can introduce seasonal graphs to better investigate seasonal climate changes. To the best of our knowledge, our proposed method is the first graph-based climate classification system.


\end{abstract}%

\section{Introduction}
\label{sec:intro}
A climate classification is an approach that is used to identify and clarify the climate characteristics between various geographic areas. It assists us in understanding 
the world's climates precisely, and cluster geographic regions based on similar climatic characteristics. 
It allows us in understanding the influence of climate on distributions of matters concerning to human’s life. Such climate classification is important in developing the solar and wind energy-related strategies and policies on regional water management planning. Gaitani \textit{et al.} \citeyearpar{Gaitani2006} used it to classify the areas of Greece with respect to the amount of received solar energy for school buildings. Climate classification is also useful in identifying the appropriate areas for farming and type of animals inhabiting in that area. The zoning of Fars province in terms of rain-fed winter wheat has been performed by Tavanpour and Ghaemi \citeyearpar{Tavanpour2016}. Climate classification has far-reaching impacts in healthcare too. The effect of climate classification in assessing quality-of-care across hospitals has been investigated by Boland \textit{et al.} \citeyearpar{Boland2017}. Such climate classification task is useful during climate-related natural hazards. Chhetri \textit{et al.} \citeyearpar{Thapa2017} have taken advantage of climate classification for analysis and standardization of the urban heat island and urban dry island effects in Matsuyama. They indicated the importance of local climate zone classification as an initial step to improve the interpretation of urban meteorological studies.

\subsection{Related work}

Since climate is a continuous phenomenon, it is not easy to classify it. However, 
this subject has attracted the attention of many researchers since a long time and has led to the development of multiple climate classification approaches. The first scheme to classify the world climates was presented by the German-Russian scientist Wladimir K\"oppen (1846-1940) in 1918 \citep{Koppen1918}. This method has been further modified by K\"oppen and Geiger \citeyearpar{Koppen1930}. The K\"oppen-Geiger climate classification system is still one of the most widely used methods. Today, this system and methods like De Martonne climate classification \citep{DeMartonne1926} are known as classical climate classifications. The classical methods are described by Essenwanger \citeyearpar{Essenwanger1986}. They group each climate type according to one or more aspects of the climate system \textit{viz.} natural vegetation, 
annual averages of temperature and the amount of precipitation. These approaches are simple but they suffer from several 
limitations. The number of variables they used is limited, while we know that each climate variable represents a distinct pattern of change and must be considered in the analysis. Also, when these methods are used on a regional scale, their accuracy depends on the quality of the data and the number of stations used in that area \citep{Raziei2017}.

Due to the limitations of classical methods and the large volume of available climate data, researchers have attempted to find alternative methods to classify the climate. Many researchers have developed new climate classification methods based on factor analysis and clustering algorithms for different areas of the world. The cluster analysis involves the grouping of data points into two (or more) different clusters. Given a set of data points, a clustering algorithm is used to classify each data point into a specific group/cluster so that the data points in the same group/cluster have similar properties and data points in different groups have highly dissimilar properties~\citep{jain2020clustering}.
These groups can then be analyzed in details to gain further knowledge about common features in each group of climate sub-regions. Therefore, cluster analysis in climatology is used to define classes of synoptic types or climate regimes \citep{Turkes2011}.

Ahmed \citeyearpar{Ahmed1997} classified the climate of Saudi Arabia using the multivariate factor-cluster analysis and showed the ability of this method to determine the climate region's boundaries. Bower \textit{et al.} \citeyearpar{Donna2007} proposed a new spatial synoptic classification scheme for western Europe. They claimed their proposed method is a generic classification method which provides a valuable tool for understanding the spatial-temporal climate variation across western Europe. Yang \textit{et al.} \citeyearpar{Yang2010} applied the factor analysis and clustering methods for grouping the areas around the Pearl River in China. In 2011, researchers suggested a hybrid clustering algorithm based on the spectral clustering and K-means method for grouping features of annual precipitation total of 96 stations in Turkey \citep{Turkes2011}. The results of their proposed method determined eight clusters of precipitation coherent zones in Turkey for the period of 1929-2007. Later in 2013, researchers attempted to redefine climate regions of Turkey by hierarchical clustering method using almost all usable climate data over the country. 
They claimed that their work could be considered as a reference for other climate-related studies of Turkey, and could be useful for the detection of Turkish climate regions \citep{Iyigun2013}.  In 2020, climate classification of China was performed by Shi and Yang \citeyearpar{Shi2020}. They introduced a spectral clustering-based method to partition 661 meteorological stations in China into several zones. The results showed that their proposed system was feasible and effective in identifying the various climatic regions.

Due to the increased availability of climate data from world-wide networks of stations and higher computing power, 
some studies \citep[e.g.,][]{Raziei2018, manandhar2018systematic} used the principal component analysis (PCA) to reduce the number of features in climate data and select the most effective variables. In climate classification field, Newnham \citeyearpar{Newnham1968} employed $19$ climate variables extracted from 70 stations in British Columbia. Newnham classified the climate using PCA and investigated its effect on tree species distribution. White and Perry \citeyearpar{White1989} described the climate classification of England and Wales based on agro-climatic data. They used PCA and clustering methods to classify climate. Perdinan and Winkler \citeyearpar{Perdinan2015} took advantage of a combination of PCA and cluster analysis to group climate stations in the USA into a small number of climate regions. They considered the minimum and maximum temperature and precipitation values to perform classification. 
Although PCA technique brings forward most important variables and captures maximum information about the climate data, it misses some information as compared to the original list of variables~\citep{alskaif2020systematic}. Furthermore, principal components are not as interpretable as original features, because the transformed basis functions are 
linear combinations of original feature space. Therefore, we seek to provide a way to be able to provide as many climate variables as possible in the analysis without changing their modality. In fact, the output of every climate classification approach depends on the selected variables.  
The results can vary for an area depending on the selected variables. Hence, it is important to develop a method 
that is able to classify regions with more climate variables, thereby generating more reliable results. 

These clustering-based classifications of climate researches employ 
generic algorithms without considering the specificity of climate data. 
The readers are referred to the paper written by Netzel \citeyearpar{Netzel2016} for further study and comparison of clustering-based climate classifications approaches. In addition to difference in techniques, local climate classification studies have been conducted for different regions of the world. Some studies have also investigated the climate of Iran and its provinces. Eslahi \citeyearpar{Eslahi2003} attempted to identify Iran climate changes by using cluster analysis. Masoodian \citeyearpar{Masoodian2003} using 27 annual climate variables, showed that climate of Iran is made by six factors and has 15 climate zones. Hatami and Khoshhal \citeyearpar{Hatami2010} classified the climate of Fars province in southern Iran, by applying the factor analysis to data gathered from weather stations and identified four different climates for Fars. Raziei \citeyearpar{Raziei2017} classified the climate of Iran via K\"oppen-Geiger system and investigated its changes during the $20^{th}$ century. The results showed Iran consisted of 9 climate types out of 31 possible K\"oppen-Geiger climate types. Furthermore, some researchers applied climate classification to a combination of historical climate data and model predictions (i.e., general circulation models (GCM)) to illustrate climate shifts over the longer time periods (e.g., Rahimi et. al using De Martonne classification system \citeyearpar{Rahimi2013} and Raziei using K\"oppen-Gieger method \citeyearpar{Raziei2017a}).

Although these climate classification techniques provide a systematic framework to classify climatic regions, however, they suffer from several shortcomings. These drawbacks include 
limitation on the number of climate variables, ignoring the specificity of climate data, and missing vital information due to the use of PCA. 
In this paper, we present a novel method that is able to use the specificity of climate data for classifying climates. We propose a 
well suited tool that can be efficiently used for local climate classification, and also eliminates the shortcomings in prior works. In this paper, we borrow techniques from graph partitioning approaches, and integrate it with 
clustering methods for an efficient climate classification. Recently, computer scientists have frequently used graphs as abstraction in modeling an application problem\citep{Donges2010, Steinhaeuser2011a, Dyn2014}. 
We model the task of climate classification as a graph partitioning problem, wherein the entire climate data is represented in the form of a graph, and then subsequently partitioned into smaller sub-graphs, with each having specific properties. 

\subsection{Contributions of our paper}
In this work, we present a new graph-based approach called Graph Partition Based Method (GPBM), to classify the regional climate accurately. 
In this paper, we benchmark our proposed method with two methods -- 
Mean Value Based Method (MVBM) and the Pearson Correlation Based Method (PCBM). 
Unlike prior works in climate classification task that represent local climate as a vector of long term monthly mean of climate variables and do not take into account the nature of climate data \citep{Netzel2016}, in graph-based method, the spatiotemporal property of climate data is preserved due to the representation type. Furthermore, due to the seasonal effects of climate variables on the climate change of regions, we introduce the seasonal graphs for the first time that 
investigate seasonal variations of regional climate. 

More details are provided in the following sections of the article. The properties of study area are explained in Section~\ref{sec:area}. In Section~\ref{sec:method}, the characteristics of data and the proposed methods are described in detail. Section~\ref{sec:res} explains the technical implementation and discusses the outputs and compares results with previous approaches. Finally, Section~\ref{sec:conc} concludes the paper and proposes the future works.

\section{The study area}
\label{sec:area}
Our study site is Fars province located in the south of Iran. This province ranges in latitude from \ang{27}N to \ang{31}N and ranges in longitude from \ang{50}E to \ang{55}E. Its total area is about $122,600$ km$^{2}$. According to latest divisions, it contains $29$ cities, wherein some of them are equipped with a synoptic station. Since the accuracy of climate classification on a regional scale depends on the quality of the data and the numbers and densities of stations used in that area \citep{Raziei2017}, we have selected an adequate numbers of synoptic stations that are uniformly distributed in the province.

\begin{figure}[hbt]
\centering
\includegraphics[width=0.8\textwidth]{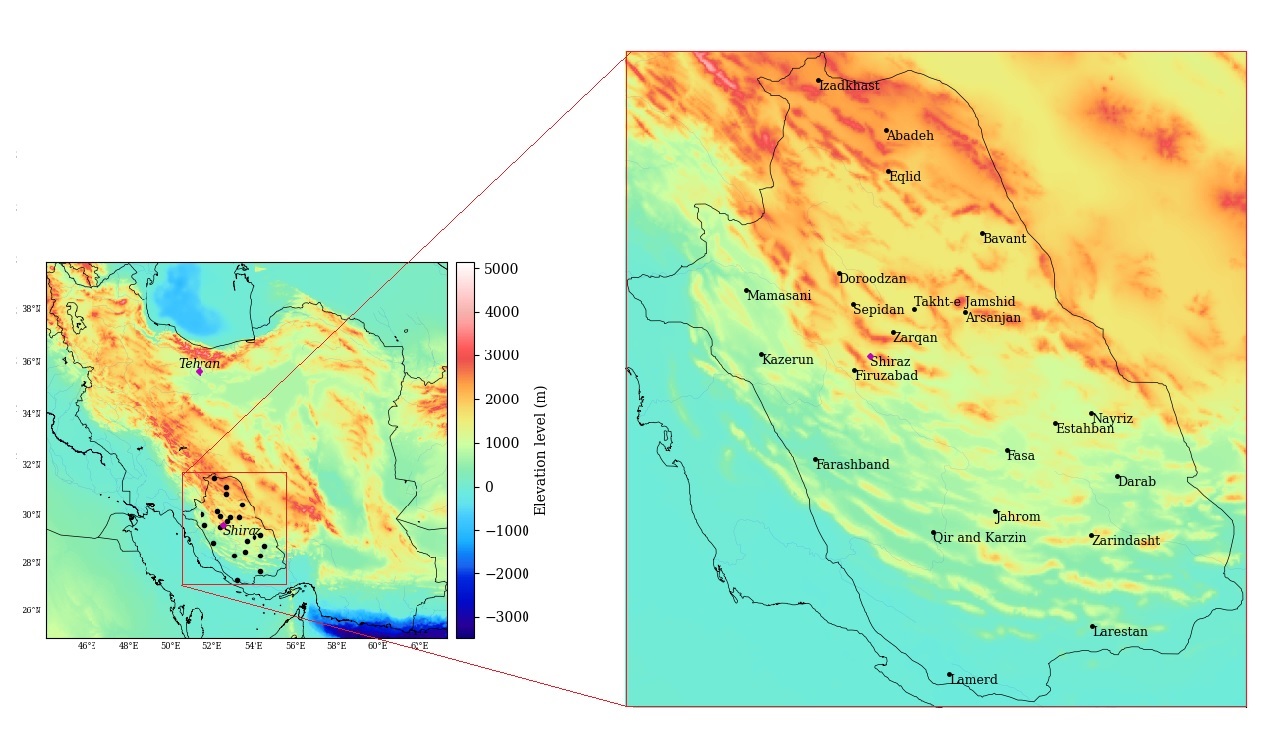}
\caption{The relative position of Fars province in Iran and the spatial distribution of synoptic stations over Fars used in the study. The map was plotted using NOAA global relief data (ETOPO1). ETOPO1 is a 1 arc-minute global relief model of Earth’s surface that integrates land topography and ocean bathymetry \citep{Etopo2009}.}
\label{fig:fars_stations}
\end{figure}

Figure~\ref{fig:fars_stations} demonstrates the relative position of Fars province in Iran and its synoptic stations. The capital of Fars is Shiraz city. According to the measurements of 2017, the average temperature of Shiraz is \ang{19.3}C, and it varies between \ang{-6}C and \ang{41}C. The climate of Fars varies through its many zones. It can be divided into three distinct climatic regions. The first zone is the mountain area of the north and northwest with moderate cold winters and partially mild summers. This is because the Zagros mountains stretches throughout the northwest. The second zone is the central regions, dominated by the Mediterranean climate 
with relatively rainy mild winters and hot dry summers. Finally, the third zone comprises the 
south and southeast zones, with cold winters and very hot summers \citep{Tavanpour2016}. There is also a wide variation in the height in the third zone. It is $3915$m above sea level in the north and $115$m in the south. The spatial variation between mean annual temperature and precipitation is high. The spatial difference in mean annual temperature is about \ang{10}C and the amount of mean annual precipitation is below 200 mm in the south and above 1000 mm in the northwest of Fars~\citep{abolverdi2016spatial}.

This province has been chosen as the study area because it is a region with high climate variability. Our review of the literature reveals that recently no effort except Hatami and Khoshhal \citeyearpar{Hatami2010} has been done to classify the climate types of this province using synoptic data. Also, as agriculture is one of the main source of income in this province, the identification of climate classification in this region is important for long-term planning and decision making on resource allocation.

\section{Climate Classification}
\label{sec:method}
In this paper, the climate types of Fars province is determined by applying a novel graph-based method. We use the monthly measured data from the selected synoptic stations in this study. 

\subsection{Data description}
\label{subsec:data}
The climate data of $24$ synoptic stations at province are provided by the Iran Meteorological Organization (\url{https://www.irimo.ir}). The longest time period of available data is 67 years from 1951 to 2017, corresponding to Shiraz station. 
The shortest one is from 2010, corresponding to Qir and Karzin station. The geographic specifications of stations are listed in the Table \ref{tab:fars_stations}.

\begin{table*}[htb]
\begin{threeparttable}
\centering
\caption{The geographic information of Fars stations.}
\label{tab:fars_stations}
\arrayrulecolor[rgb]{0.788,0.788,0.788}
\begin{tabular}{llll!{\color{black}\vrule}llll} 
\arrayrulecolor{black}\hline
Station & Longitude & Latitude & Height\tnote{*} (m) & Station & Longitude & Latitude & Height(m) \\ 
\hhline{>{\arrayrulecolor[rgb]{0.788,0.788,0.788}}--->{\arrayrulecolor[rgb]{0.788,0.788,0.788}}-|>{\arrayrulecolor[rgb]{0.788,0.788,0.788}}----}
\rowcolor[rgb]{0.929,0.929,0.929} Abadeh & 52$^{\circ}$ 40’ & 31$^{\circ}$ 11’ & 2030 & Sepidan & 52$^{\circ}$ & 30$^{\circ}$ 14’ & 2201 \\ 
\cline{1-3}\arrayrulecolor[rgb]{0.788,0.788,0.788}\cline{4-4}\arrayrulecolor[rgb]{0.788,0.788,0.788}\cline{5-8}
Arsanjan & 53$^{\circ}$ 16’ & 29$^{\circ}$ 56’ & 1703 & Estahban & 54$^{\circ}$ 2’ & 29$^{\circ}$ 5’ & 1690 \\ 
\hhline{--->{\arrayrulecolor[rgb]{0.788,0.788,0.788}}-|>{\arrayrulecolor[rgb]{0.788,0.788,0.788}}----}
\rowcolor[rgb]{0.929,0.929,0.929} Eqlid & 52$^{\circ}$ 38’ & 30$^{\circ}$ 54’ & 2300 & Izadkhast & 52$^{\circ}$ 7’ & 31$^{\circ}$ 32’ & 2188 \\ 
\cline{1-3}\arrayrulecolor[rgb]{0.788,0.788,0.788}\cline{4-4}\arrayrulecolor[rgb]{0.788,0.788,0.788}\cline{5-8}
Bavanat & 53$^{\circ}$ 40’ & 30$^{\circ}$ 28’ & 2231 & Takht-e Jamshid & 52$^{\circ}$ 54’ & 29$^{\circ}$ 56’ & 1605 \\ 
\hhline{--->{\arrayrulecolor[rgb]{0.788,0.788,0.788}}-|>{\arrayrulecolor[rgb]{0.788,0.788,0.788}}----}
\rowcolor[rgb]{0.929,0.929,0.929} Jahrom & 53$^{\circ}$ 32’ & 28$^{\circ}$ 29’ & 1082 & Darab & 54$^{\circ}$ 17’ & 28$^{\circ}$ 47’ & 1098 \\ 
\cline{1-3}\arrayrulecolor[rgb]{0.788,0.788,0.788}\cline{4-4}\arrayrulecolor[rgb]{0.788,0.788,0.788}\cline{5-8}
Doroodzan & 52$^{\circ}$ 17’ & 30$^{\circ}$ 11’ & 1650 & Zarqan & 52$^{\circ}$ 43’ & 29$^{\circ}$ 47’ & 1596 \\ 
\hhline{--->{\arrayrulecolor[rgb]{0.788,0.788,0.788}}-|>{\arrayrulecolor[rgb]{0.788,0.788,0.788}}----}
\rowcolor[rgb]{0.929,0.929,0.929} Zarrin Dasht & 54$^{\circ}$ 25’ & 28$^{\circ}$ 21’ & 1029 & Shiraz & 52$^{\circ}$ 36’ & 29$^{\circ}$ 32’ & 1484 \\ 
\cline{1-3}\arrayrulecolor[rgb]{0.788,0.788,0.788}\cline{4-4}\arrayrulecolor[rgb]{0.788,0.788,0.788}\cline{5-8}
Safashahr & 53$^{\circ}$ 9’ & 30$^{\circ}$ 35’ & 2251 & Fasa & 53$^{\circ}$ 41’ & 28$^{\circ}$ 58’ & 1288 \\ 
\hhline{--->{\arrayrulecolor[rgb]{0.788,0.788,0.788}}-|>{\arrayrulecolor[rgb]{0.788,0.788,0.788}}----}
\rowcolor[rgb]{0.929,0.929,0.929} Firuzabad & 52$^{\circ}$ 33’ & 28$^{\circ}$ 53’ & 1362 & Farashband & 52$^{\circ}$ 6’ & 28$^{\circ}$ 48’ & 782 \\ 
\cline{1-3}\arrayrulecolor[rgb]{0.788,0.788,0.788}\cline{4-4}\arrayrulecolor[rgb]{0.788,0.788,0.788}\cline{5-8}
Kazerun & 51$^{\circ}$ 39’ & 29$^{\circ}$ 36’ & 860 & Qir and Karzin & 53$^{\circ}$ 3’ & 28$^{\circ}$ 28’ & 746 \\ 
\hhline{--->{\arrayrulecolor[rgb]{0.788,0.788,0.788}}-|>{\arrayrulecolor[rgb]{0.788,0.788,0.788}}----}
\rowcolor[rgb]{0.929,0.929,0.929} Lamerd & 53$^{\circ}$ 12’ & 27$^{\circ}$ 22’ & 405 & Larestan & 54$^{\circ}$ 17’ & 27$^{\circ}$ 42’ & 792 \\ 
\cline{1-3}\arrayrulecolor[rgb]{0.788,0.788,0.788}\cline{4-4}\arrayrulecolor[rgb]{0.788,0.788,0.788}\cline{5-8}
Nayriz & 54$^{\circ}$ 20’ & 29$^{\circ}$ 12’ & 1632 & Mamasani & 51$^{\circ}$ 32’ & 30$^{\circ}$ 4’ & 972 \\
\cline{1-3}\arrayrulecolor[rgb]{0.788,0.788,0.788}\cline{4-4}\arrayrulecolor[rgb]{0.788,0.788,0.788}\cline{5-8}
\arrayrulecolor{black}\hline
\end{tabular}
\begin{tablenotes}
\item {*It is the height from the see level (i.e. Elevation level)}
\end{tablenotes}
\end{threeparttable}
\end{table*}

The data set contains the following sensor measurements: maximum temperature, minimum temperature, average temperature, total precipitation, average wind speed, direction and maximum wind speed. In the downloaded data set, 
$414$ data points are empty. 
We filled the empty data points by the average monthly values for each month of the year. Therefore, 
we have a data matrix with $8$ columns including seven climate variables and one time span column for each stations. In other words, there is a matrix of time series for each station. The largest matrix with $804$ rows and $8$ columns belongs to Shiraz, and the smallest one belongs to Qir and Karzin with $96$ rows and $8$ columns. We note here that one of the stations named Fasa was excluded from the data set since the amount of total precipitation was not available for this station. 

\begin{figure*}[htb]
  \begin{center}
    \subfloat[Average temperature]{\includegraphics[width=0.3\textwidth]{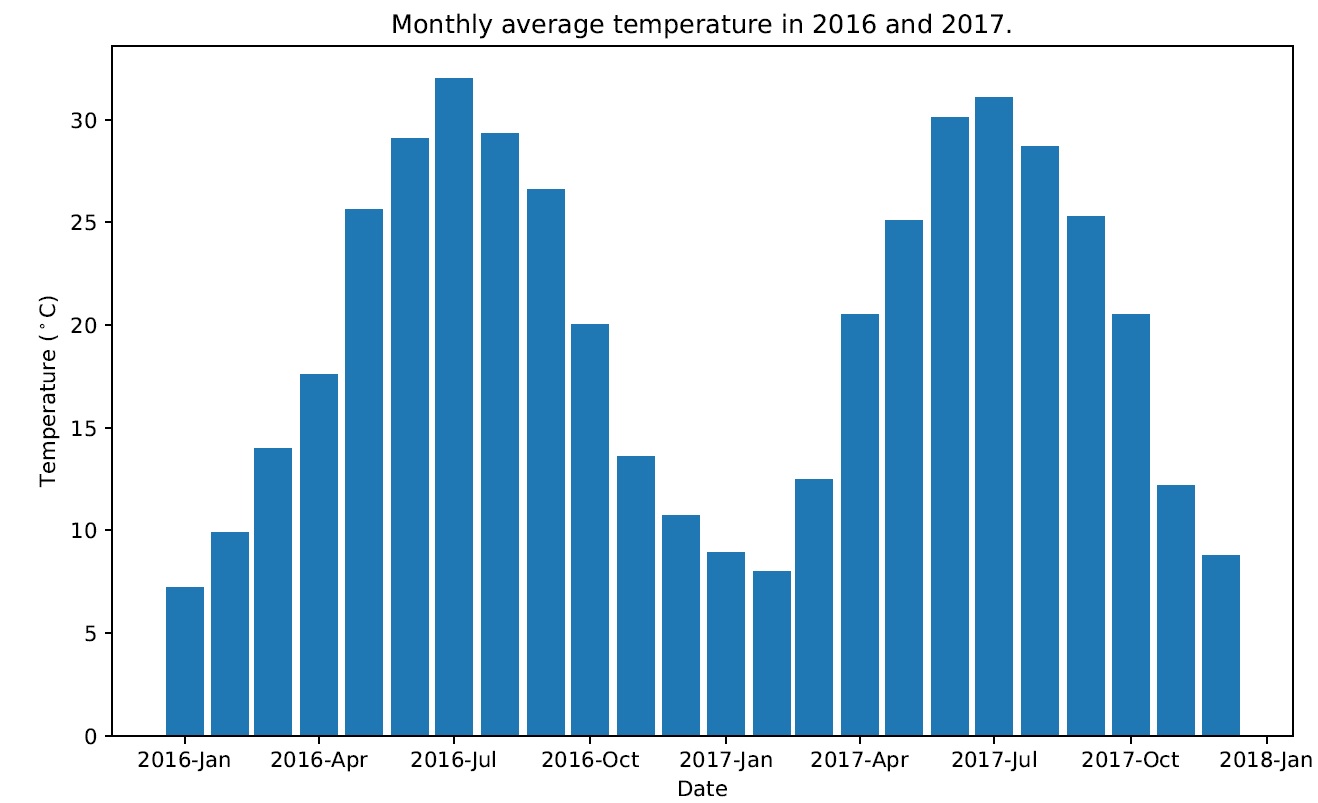}}
    \subfloat[Minimum temperature]{\includegraphics[width=0.3\textwidth]{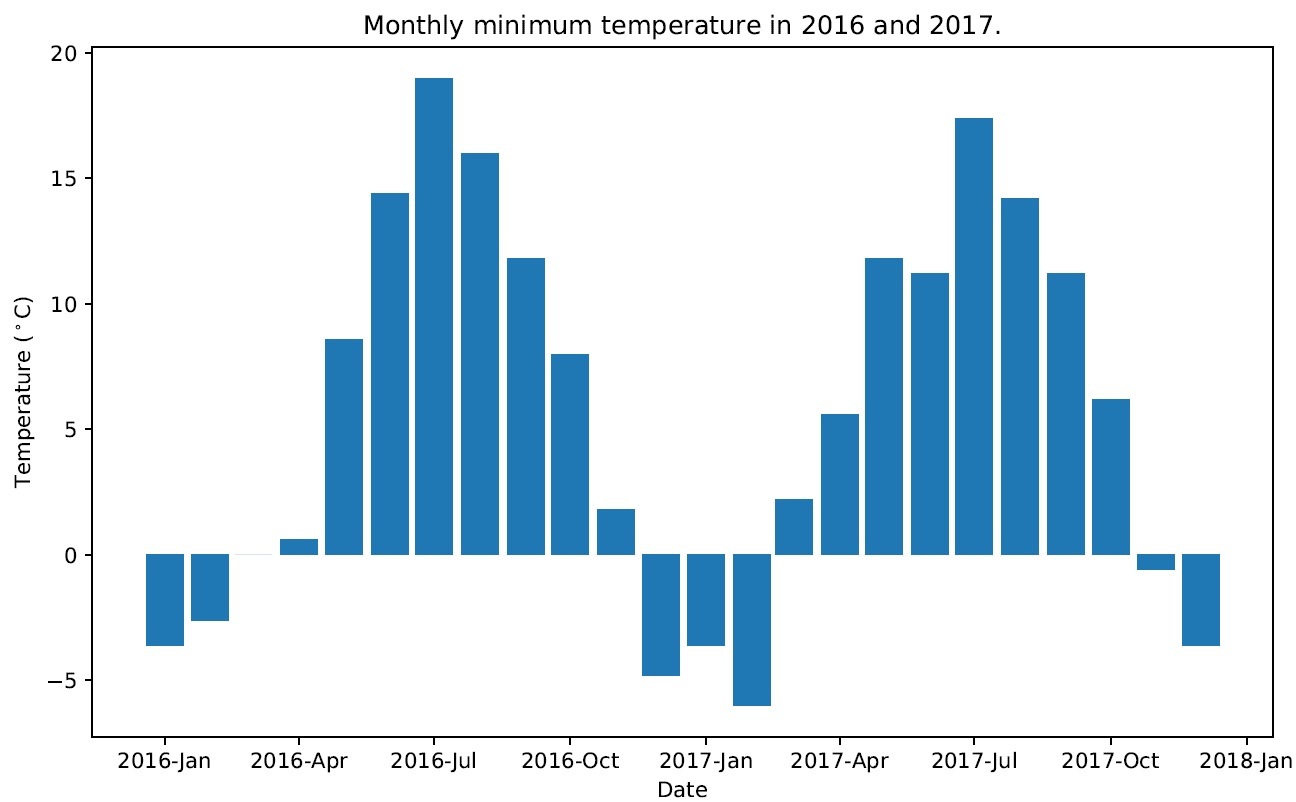}}
    \subfloat[Maximum temperature]{\includegraphics[width=0.3\textwidth]{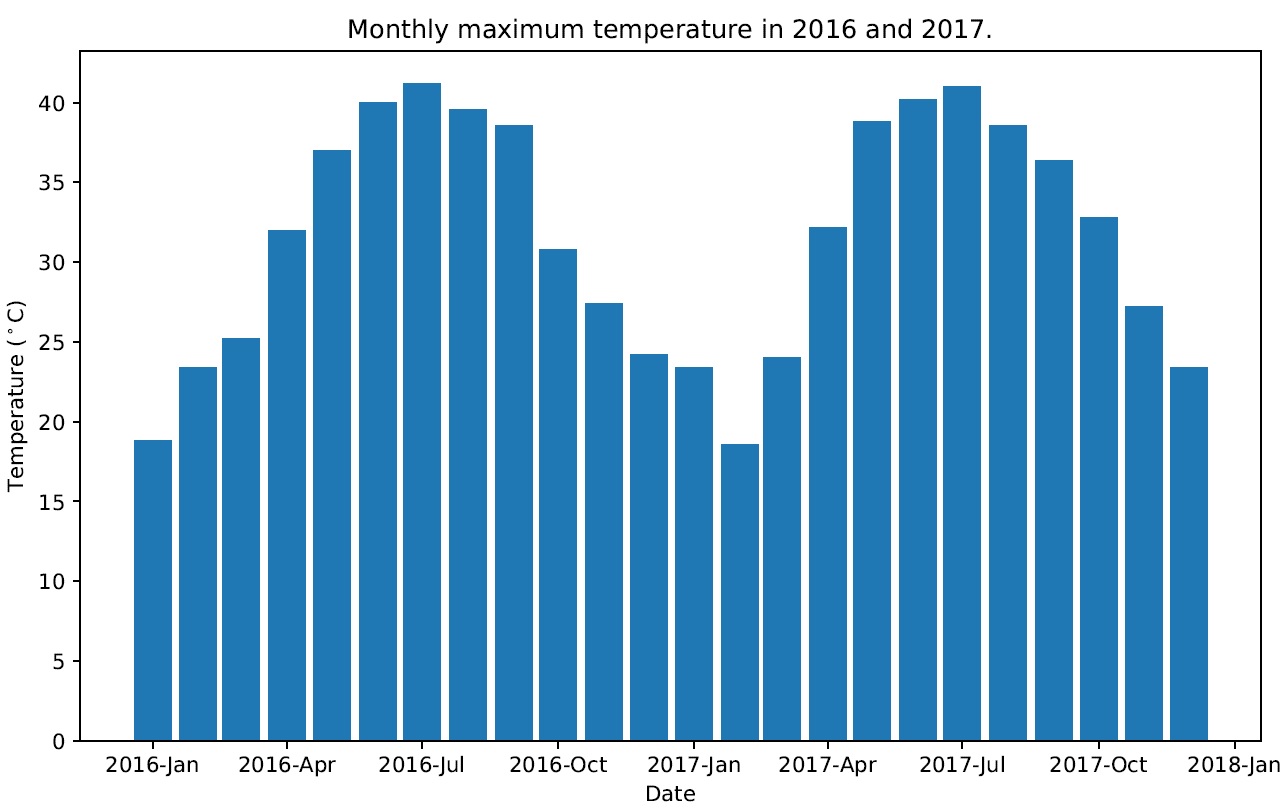}}\\
    \subfloat[Wind speed]{\includegraphics[width=0.25\textwidth]{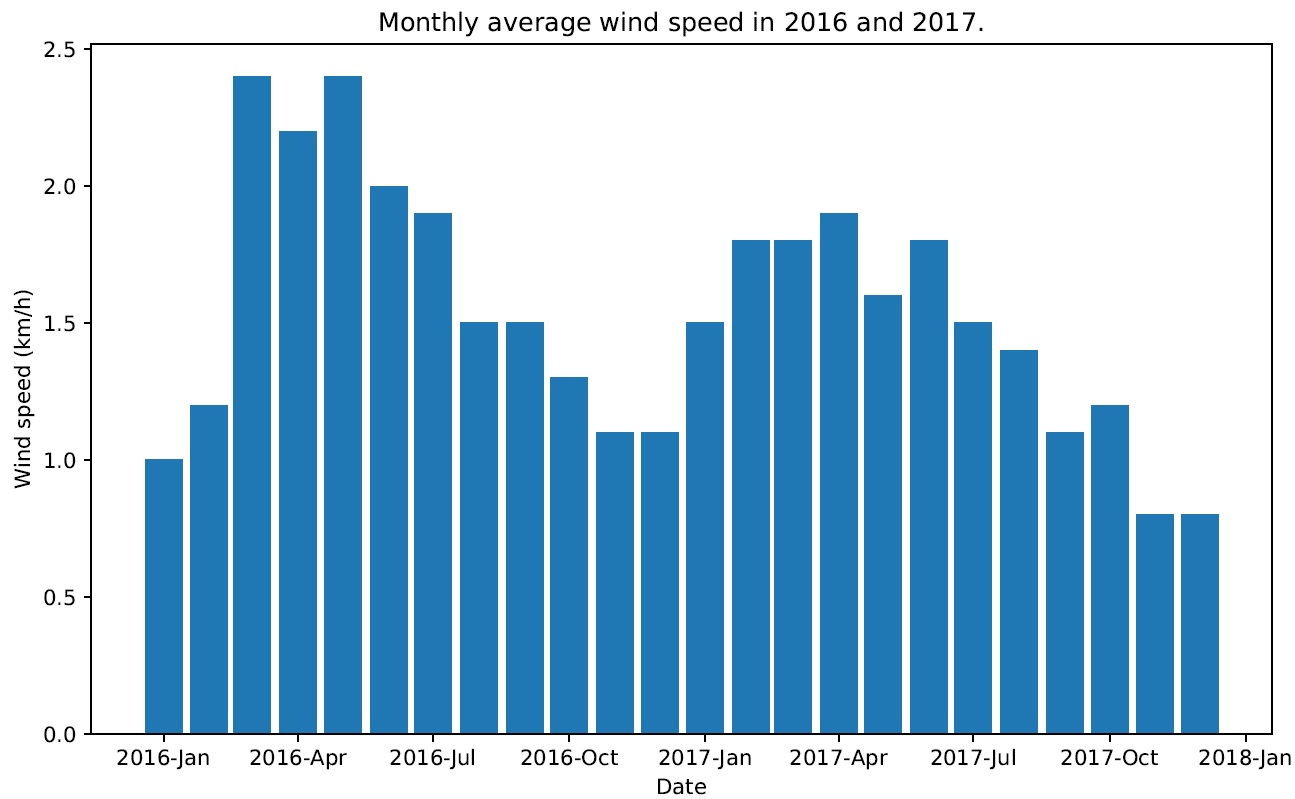}}
    \subfloat[Maximum wind]{\includegraphics[width=0.25\textwidth]{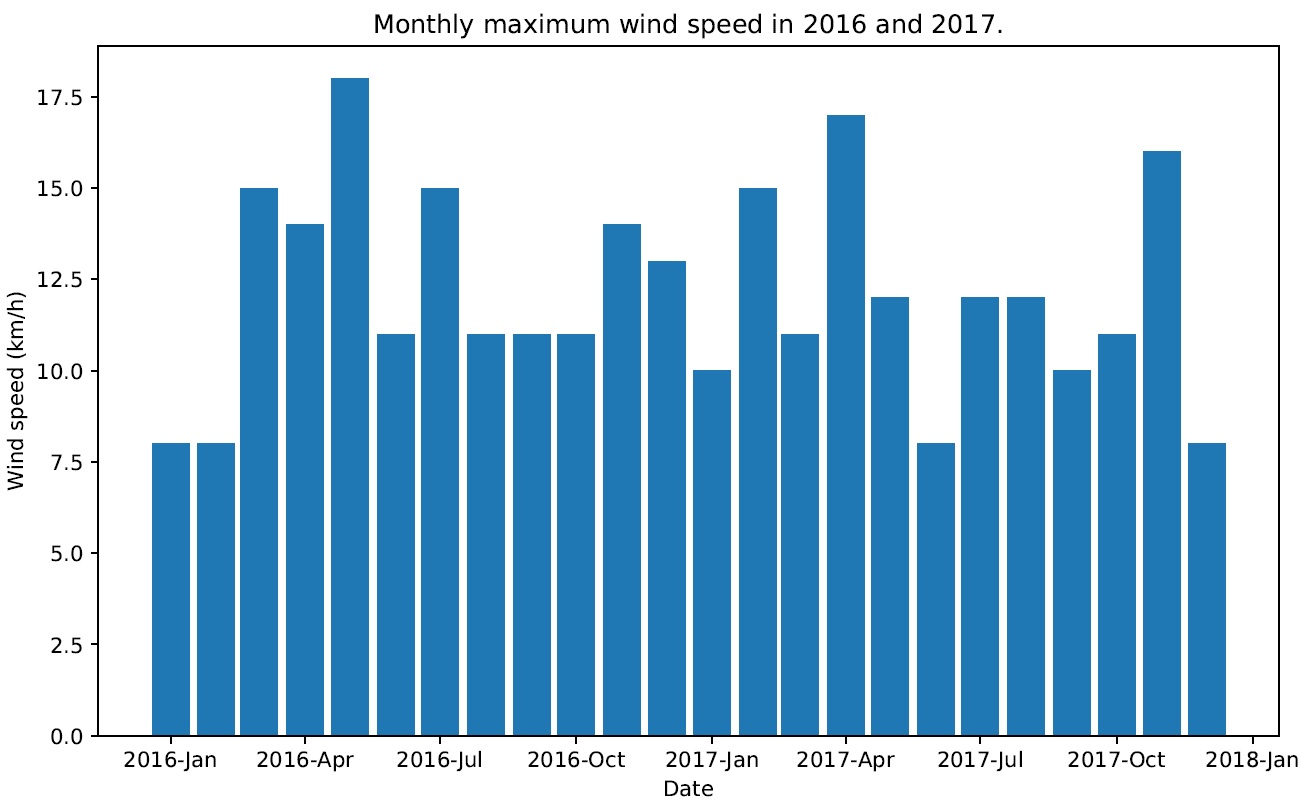}}
    \subfloat[Wind direction]{\includegraphics[width=0.25\textwidth]{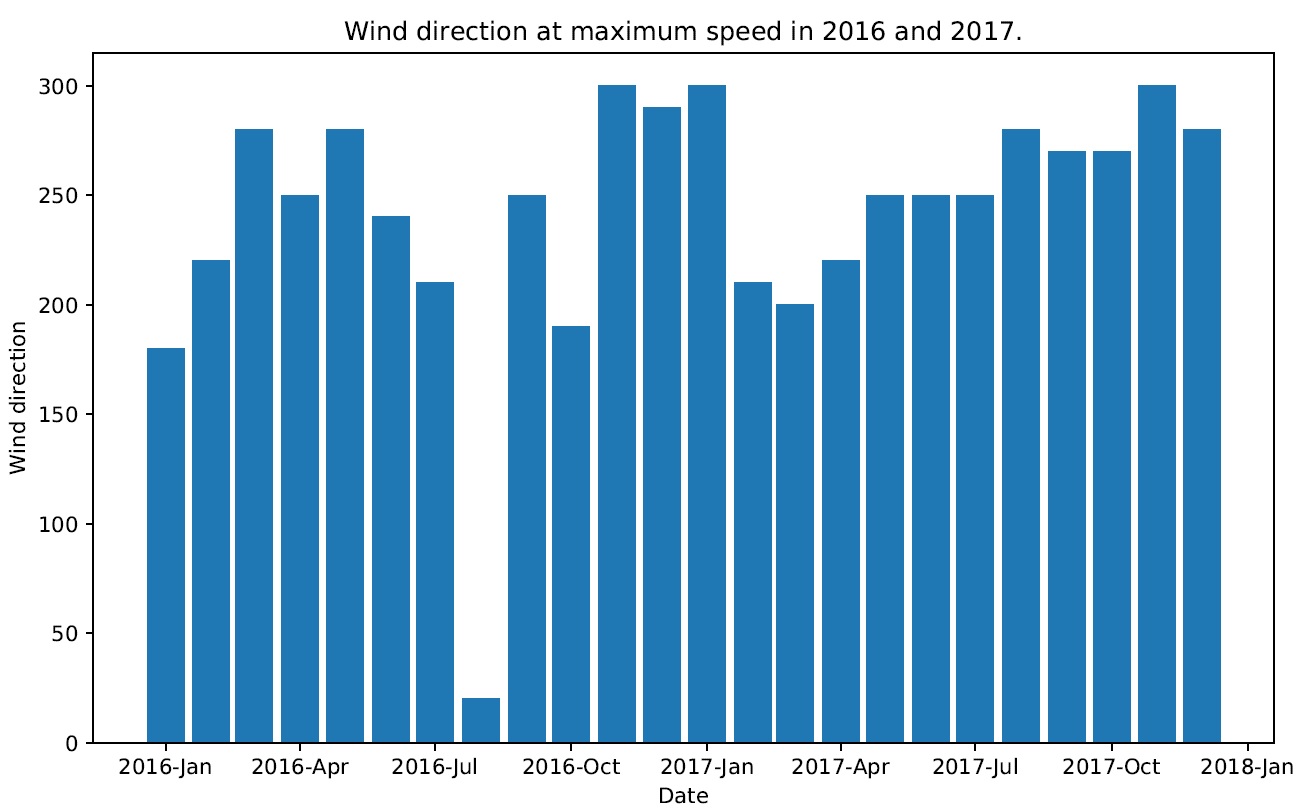}}
    \subfloat[Total precipitation]{\includegraphics[width=0.25\textwidth]{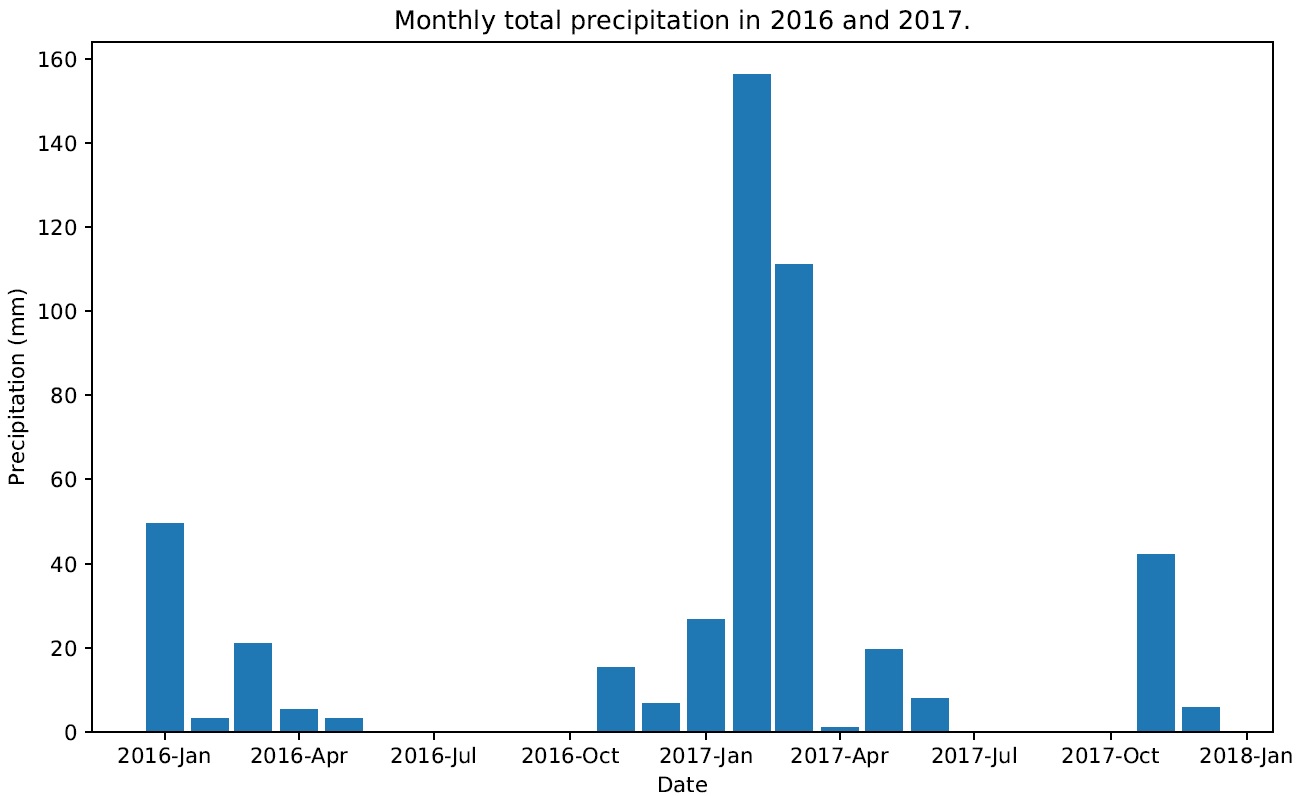}}
  \end{center}
  \caption{Climate observation of Shiraz station in 2016 and 2017.}
  \label{fig:shz_observation}
\end{figure*}

Figure~\ref{fig:shz_observation} shows a representative time series for each variable of the Shiraz station, in order to have a better understanding of our dataset. It is important for us to decompose the time series into several components to explore its underlying pattern. We decompose the data into the trend and seasonality component~\citep{dev2018solar}. 
The trend indicates an increase or decrease in the temperature value, whereas the seasonality refers to the repeating short-term cyclic nature in the temperature data. The trend and seasonality can be observed intuitively in the charts. The seasonal pattern is seen in the temperature variable (average, maximum and minimum) where the pattern is repeated annually. Also, we use the additive decomposition technique to estimate the seasonal components and determine the trend of time series 
\citep{Hyndman2018}.

\begin{figure*}[htb]
  \begin{center}
    \subfloat[Original time series]{\includegraphics[width=0.33\textwidth]{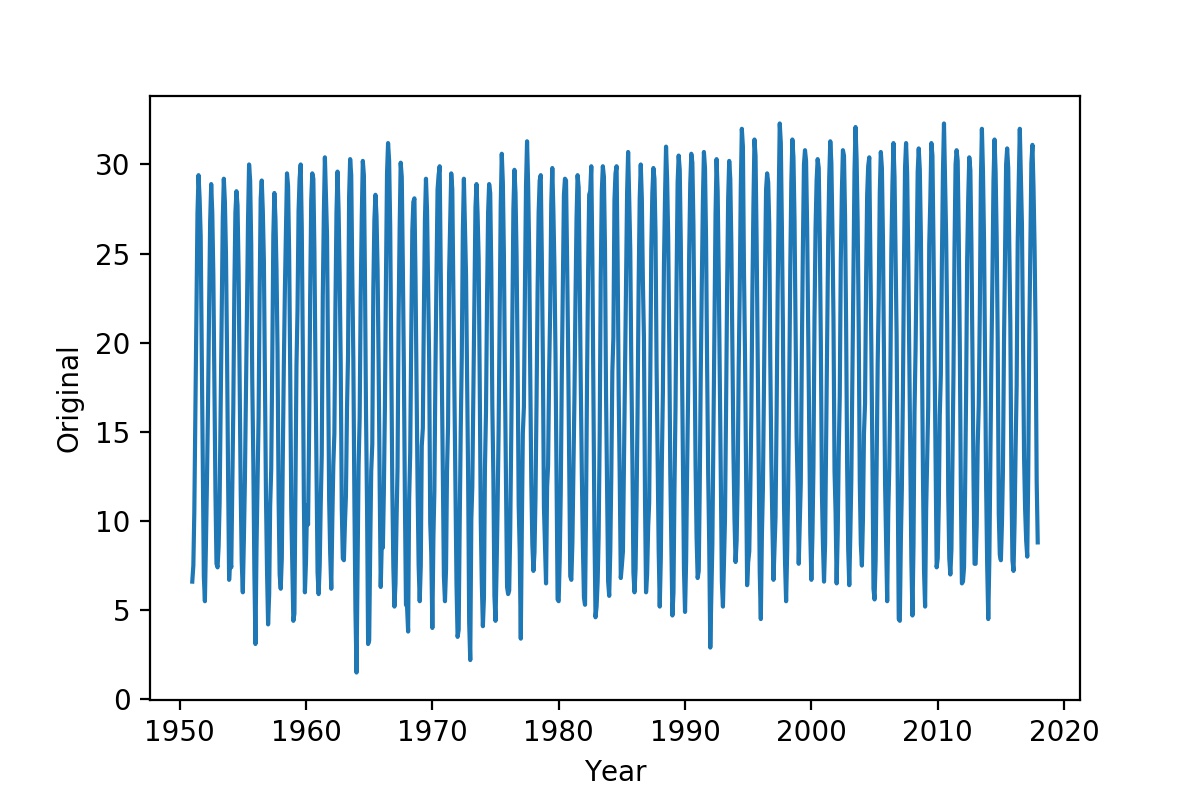}}
    \subfloat[Trend]{\includegraphics[width=0.33\textwidth]{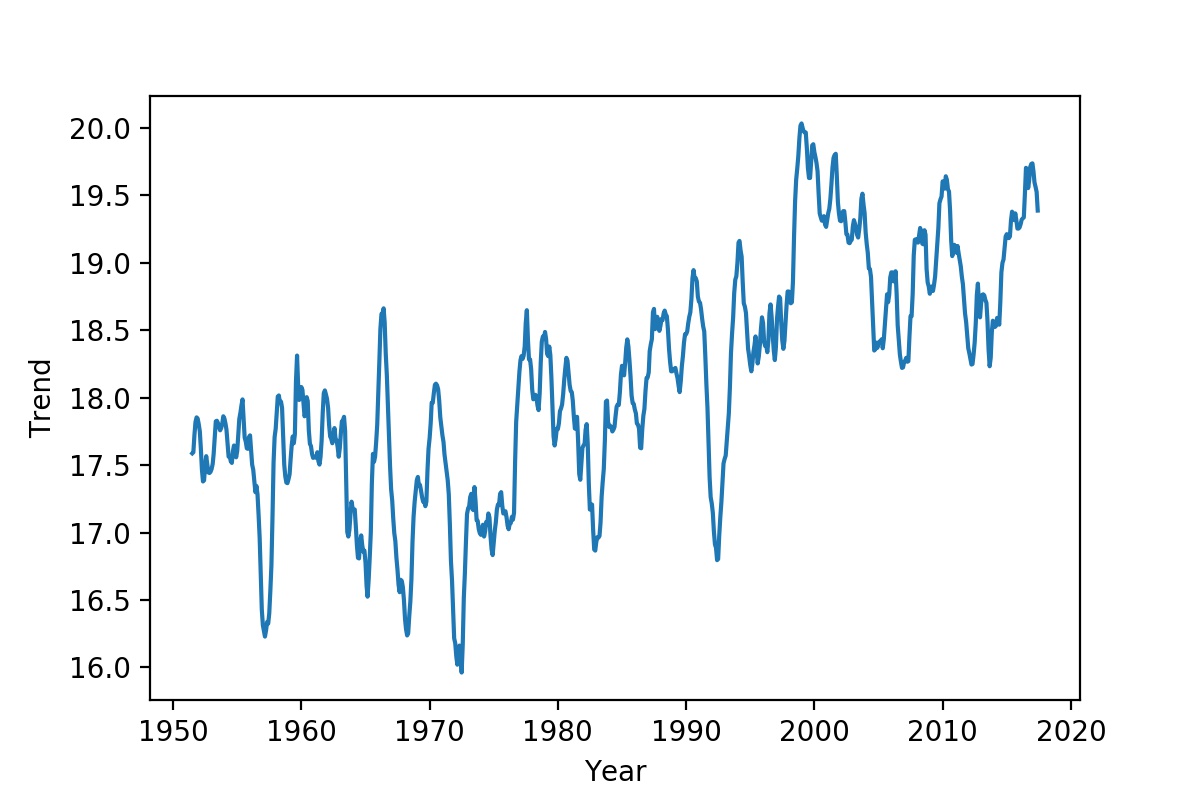}}
    \subfloat[Seasonality]{\includegraphics[width=0.33\textwidth]{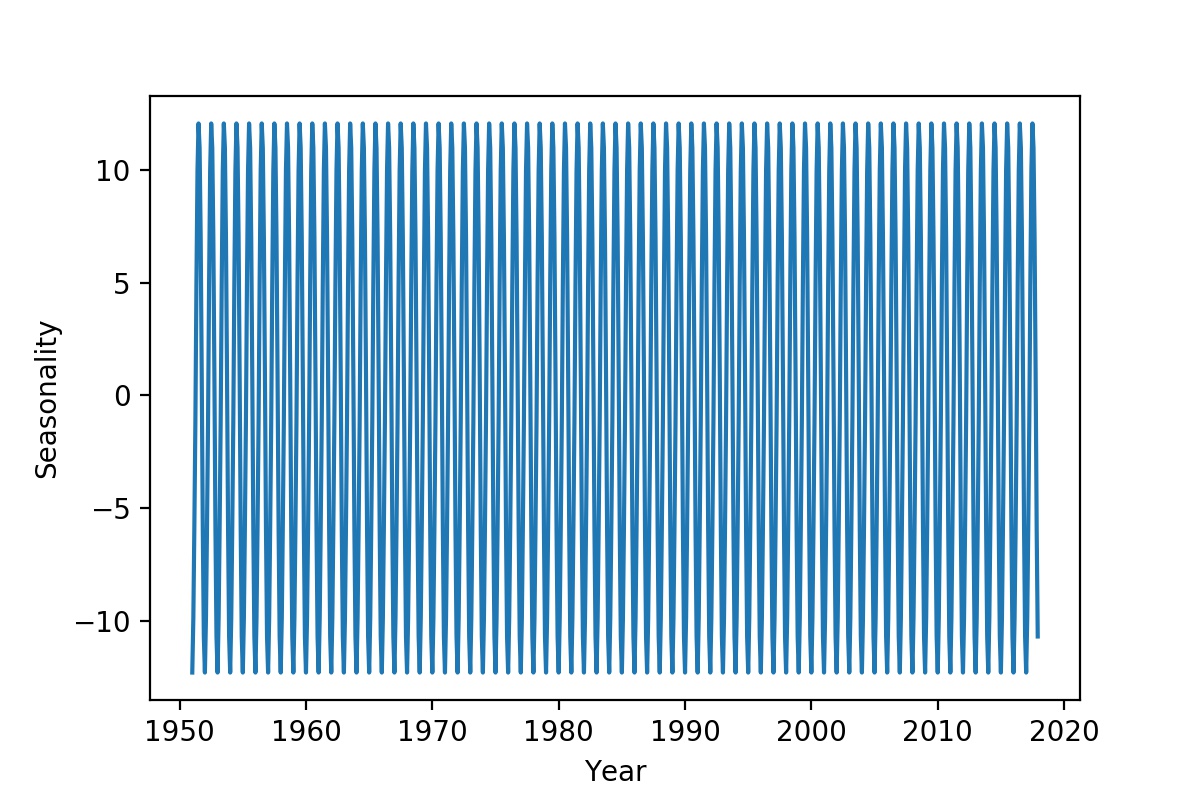}}
  \end{center}
  \caption{The decomposition components of Shiraz average temperature time series.}
  \label{fig:decompsTS}
\end{figure*}

We apply the additive model to average temperature and illustrate its components at Fig.~\ref{fig:decompsTS}. The trend chart shows the clear uptrend in mean temperature in Shiraz over $67$ years. In the wind direction (cf.\ Fig.\ 2f), it may not be possible to take into account the seasonal pattern. There is also a seasonal pattern for the total precipitation (cf.\ Fig.\ 2g). However, due to heavy rainfall occurring in a few months, its intensity is greater than that of other variables. 
Such decomposition leads us to an improved understanding of the input temperature data of Shiraz.

\subsection{Proposed method}
\label{subsec:method}
In this section, we describe the proposed method for climate classification. We also provide a primer of the 
clustering algorithm that is used in our proposed method. 

There are many clustering algorithms in literature. One of the most common and simplest algorithms that researchers have applied to climate is K-means \citep{macqueen1967}. It is one of the simplest algorithms to solve the clustering problem. The algorithm needs to pre-define the number of clusters (i.e. the number of centroids of clusters) as `K'. It iteratively calculates the closeness between samples to each centroid and generates the compact groups of them as clusters. The k-means attempts to construct a circular-shaped clusters around the centroids. Also, it does not let data points that are far away from each other to be grouped to the same cluster. 
Accordingly, K-means cannot detect non-convex and non-circular clusters~\citep{Bhargav2016}. 

Identifying clusters with different shapes is a challenge in clustering algorithms. Therefore, during climate clustering, we can observe that districts with significant distance between each other can possess similar climate. 
Therefore, it is obvious that the shape of cluster is non-circular and discontinuous. The K-means algorithm has poor performance in detecting clusters with various geometric shapes. Therefore, the shape of climate clusters (i.e. areas with similar climates) may not be circular and convex due to the topography of the regions and the existence of mountains and lakes with different ranges. Based on these drawbacks of the popular k-means algorithm, we select the spectral clustering method. It is one of the most popular clustering algorithms that uses K-means as its clustering technique, but does not make any assumptions on the shape of the clusters \citep{Donath1973, jain2021validating}. This property of spectral clustering wherein it does not make assumption on shape of clusters is necessary for the climate classification.

The spectral or subspace clustering treats the task of data clustering as a graph partitioning problem. It refers to a group of clustering methods that utilize the eigenvalues of a similarity matrix for dimensionality reduction. This similarity matrix constructed from the data is referred as the spectrum. 
They create clusters in the lower-dimensional space by partitioning the data into $k$ subsets. 
In recent years, it has become one of the most common clustering methods 
widely used by climatology researchers \citep[e.g.,][]{Fouedjio2017, Shi2020}. There are various extensions of this approach. We use the method introduced by Y. Ng \textit{et al.} \citeyearpar{YNg2001} which was shown to yield proper results on a wide variety of challenging clustering problem. The steps of the spectral clustering algorithm are briefly described in Algorithm \ref{algo:spectral}. The readers are referred to Luxburg \citeyearpar{VonLuxburg2007} for detailed discussion of the spectral clustering algorithm.

\SetKwInput{KwInput}{Input}                
\SetKwInput{KwOutput}{Output}              

\begin{algorithm}
  \KwInput{a set of $n$ points $S = \{s_{1}, ..., s_{n}\}$ in $R^n$}
  \KwOutput{k subsets including data points}
  
Pre-processing stage: \linebreak
-	Create the affinity matrix $A \in R^{n*n}$ defined by \linebreak $A_{ij} = exp(-\left \| s_{i}-s_{j} \right \|^2/2\sigma^2)$ for $i \neq j$, and $A_{ii}=0$.\linebreak
-	Construct the graph Laplacian matrix $L = D^{-1/2}AD^{-1/2}$ where D is a diagonal matrix whose elements are the degrees of the nodes of the graph $d_{ii}=\sum_{j=1}^{n}Aij$. \\
Spectral representation stage: \linebreak
-	Compute eigenvalues and eigenvectors of the Laplacian matrix $L$ \linebreak
-	Pick out $x_1, x_2, ..., x_k$ the k largest eigenvectors of L to define a k-dimensional subspace.  \linebreak
-	Form the matrix $X = [x_1x_2...x_k] \in R^{n*k}$ by stacking the eigenvectors in columns \linebreak
-	Re-normalize each row of $X$ to have unit length and consider each row as a point in lower-dimensional representation $R^k$. \\
Clustering stage: \linebreak
-	Cluster the new points into k classes using K-means algorithm. \linebreak
-   Finally, assign the original point $s_i$ to cluster $j$ if and only if row $i$ of the $X$ was assigned to cluster $j$.

\caption{The spectral clustering algorithm steps.}
\label{algo:spectral}
\end{algorithm}

The main tools of the spectral clustering are the graph Laplacian matrices. As specified in the Algorithm \ref{algo:spectral}, we compute the affinity matrix $A$ 
using Gaussian kernel. The scaling parameter $\sigma$ controls how rapidly the affinity $A_{ij}$ falls off with the distance between $s_i$ and $s_j$ \citep{YNg2001}. This parameter is set automatically. After that, the Laplacian matrix $L$ is constructed via the diagonal matrix D whose ($i$,$i$)-element is the sum of elements of A's $i$-th row. At the second step, the top $k$ eigenvectors of $L$ are chosen to be orthogonal to each other in the case of repeated eigenvalues. These $k$ eigenvectors make a $k$-dimensional projection of the points. Finally, we apply the 
standard clustering method to the new space in order to find the clusters. We use the K-means algorithm for clustering.  
   
At this point, the issue we deal with is that the optimal number of clusters $k$ is unknown. There is no general theoretical solution to find this value for any given data set. However, there are many technical approaches for estimating $k$ such as elbow by Thorndike \citeyearpar{Thorndike1953}, silhouette index by Rousseeuw \citeyearpar{ROUSSEEUW1987}, gap statistic by Tibshirani et al. \citeyearpar{Tibshirani2003} and the method proposed by Sugar and Jemes \citeyearpar{sugar2003}. In general, these methods compare the results of multiple runs with different k clusters and choose the best one according to a given criterion. However, from a purely statistical side there seems to be no reliable approach to determine the number of clusters \citep{Mahlstein2010}. 

Although in some climatological studies \citep[e.g,][]{Mahlstein2010, Raziei2018, Shi2020} researchers have used these methods to estimate the number of clusters, eventually they have chosen $k$ according to the conditions of the problem and their prior knowledge of it. The determination of proper number of clusters requires specific solutions which depend on the problem at hand \citep{Mahlstein2010}. Therefore, we need to be careful in the choice of the value $k$. This is because increasing $k$ yields in smaller error function values by definition, but also increases the risk of over-fitting.

By applying the spectral clustering algorithm and graph partitioning technique, we introduce a model for climate classification. We refer this framework as GPBM: \textbf{G}raph \textbf{P}artitioning \textbf{B}ased \textbf{M}ethod in the subsequent discussions of this paper. 
The steps of GPBM method are written in Algorithm~\ref{algo:GPBM}.

\SetKwInput{KwInput}{Input}                
\SetKwInput{KwOutput}{Output}              

\begin{algorithm}
\DontPrintSemicolon
  
  \KwInput{n: \# of stations, m:\# of climate variables, k:\# of clusters, $\rho$: density percentage}
  \KwOutput{A list includes the climate classification of the stations}
  \KwData{Synoptic data of n stations include m climate variables}
  Data set production: \linebreak
  - Extract the time series for each variable (i.e. $TS_{ij}  \quad i \in \{1, 2, ..., m\} \quad and \quad j \in \{1, 2, ..., n\}$) from the input files and produce m data sets, separately.\linebreak
  $DS_1 = \{TS_{11}, TS_{12}, .., TS_{1n}\}$ \linebreak
  $DS_2 = \{TS_{21}, TS_{22}, .., TS_{2n}\}$ \linebreak
  ... \linebreak
  $DS_m = \{TS_{m1}, TS_{m2}, .., TS_{mn}\}$ \\
  
  Data clustering: \linebreak
  - Call m spectral clustering algorithms with K=k on $DS_i \quad i \in \{1, 2, ..., m\}$, simultaneously. \linebreak
  - Generate $C_1$ to $C_m$ lists includes clustering results. \linebreak
  - Calculate the similarity between both station pairs ($W_{i,j}$) by formula (\ref{eq:w}). \\
  
  Graph construction: \linebreak
  - Construct undirected weighted graph G based on calculated weights ($W_{i,j}$)  \linebreak
  - Calculate density of G, called d \\
  
  Graph pruning: \linebreak
  - Run forward pruning technique with termination criteria $d'>=\rho * d$ where $d'$ is the density of new graph. \linebreak
  - Estimate the threshold, $\tau$, based on new graph for pruning G.  \linebreak
  - Remove the edges with $W < \tau$ from G to make pruned graph $G'$\\
  
  Graph partitioning: \linebreak
  - Partition $G'$ and extract zones to determine climate classification.
  
\caption{We explain the GPBM algorithm in a detailed fashion.}
\label{algo:GPBM}
\end{algorithm}

At first step, we consider only time series of one climate variable for each data sample in the form of a multi-dimensional feature vector. Subsequently, we apply the spectral clustering algorithm 
to this new space and obtain the data clusters. 

Given the above approach, for each climate variables, there is a clustering result that is obtained by running the clustering algorithm separately on its related data. Unlike the PCA method that requires data standardization before applying PCA, we do not need to standardize data because only one variable is considered in each clustering implementation. In the next step, we use these clustering results to define a similarity measure between two stations (W). It is calculated using equation (\ref{eq:w}). 

\begin{equation}
    W_{i,j}=\sum_{c=1,i\neq j}^{m} w_{ijc}
\label{eq:w}
\end{equation}

Where $w_{ijc}$ is $1$ if station $i$ and $j$ are in same cluster when the clustering algorithm for \(c^{th}\) climate variable is executed. Otherwise, it is zero. And $m$ is the number of clustering algorithms which execute separately. 
In this way, all available climate variables are exploited in the graph construction process, and not restricted to just one variable such as temperature or precipitation~\citep{Tomovic2018}.

Note there will be an edge between two nodes if at least in one clustering result, both are in a same cluster. The weight of each edge or similarity between two nodes is equal to the number of times the two stations are located in the same cluster. Hence, the minimum value of similarity is zero and the maximum value is $m$. 

Eventually, $G$ = ($V$, $E$) is a weighted undirected graph where $V$ is set of physical location of stations as nodes and $E$ is set of weighted-edges. To assign a weight to the edges, we define a simple similarity measure based on clustering results, as defined in formula \ref{eq:w}.

In summary, the graph $G$ is constructed in this way that let each station be a node, and for each pair of nodes it draws an edge and assigns it a weight equal to the clustering-based similarity measure defined above. 
A lower weight indicates more climatic difference between the two nodes. In order to further 
reduce the computational cost, our next step is pruning. For pruning, a threshold $\tau$ on weight is considered. To specify the value of $\tau$, we use a pre-pruning technique. This technique is also known as forward pruning. It first sorts the edges based on their weight, then adds the edges with highest weight to null graph 
(edges with $W$=$m$, then $W$=$m-1$ and so on) until the density of created graph will be at least 70\% ($\rho=0.7$) of density of $G$. Consequently, edges with weight of less than $\tau$ are removes and a pruned graph is created.

The spatiotemporal nature of climate data is one of the great challenges in our case. To overcome this issue, researchers often consider only a single variable by using the time series data as 
a feature vector during its application to the clustering algorithms. However in this study, we deal with a multivariate problem and our goal is to determine the climate classification with respect to all available variables. Therefore, we first cluster the data in the same way (\textit{i.e.} as separated 
feature vector for each variables), and then take advantage of the partitioning approaches in the graph theory. With this in mind, the preservation of spatiotemporal feature is a part of the nature of the algorithm. In other words, the GPBM preserves the temporal aspect of the data since it performs the clustering on the each variable as time series separately in the data clustering stage. The GPBM also preserves the spatial property of the data by considering all the stations as samples in clustering algorithm. The difference between this method with other methods is that GPBM employs all climate variables and it is not limited to using one variable in order to retain the spatiotemporal nature of the data.

The final step involves partitioning the pruned graph to determine climate classification. The objective function for partitioning is to find the groups, intra-cluster density vs.\ inter-cluster sparsity~\citep{Schulz2016}. As a result, each partition/cluster represents cities with similar climate.

\subsection{Benchmarking methods}
To evaluate the effectiveness of preserving the spatio-temporal nature of the data and the type of data representation in the GPBM method, we benchmark our proposed method with two methods based on the spectral clustering algorithm. The first method is called \textit{Mean Value Based Method} (MVBM) and the second method is called \textit{Pearson Correlation Based Method} (PCBM). These methods apply the spectral clustering algorithm to data stored in two different ways. Note that the disadvantage of these benchmarking methods is that they do not take into account the temporal nature of the data.

Since the clustering methods will often produce different results on the same data \citep{Netzel2016}, the spectral clustering algorithm (Algorithm \ref{algo:spectral}) have been used in all three methods so that their performance does not depend on the efficiency of different clustering algorithms.

\subsubsection{MVBM method}
In this approach, we transform the data matrix into a vector, by replacing each column with its average values. 
This transformation for each data sample is formulated as Equation~\ref{eq:mvbm}.
 
 \begin{equation}
 \begin{split}
 S_i = \begin{bmatrix}
v_{11} & v_{12} & ... & v_{1m}\\ 
v_{21} & v_{22} & ... & v_{2m}\\ 
... & ... & ... & ... \\ 
v_{t1} & v_{t2} & ... & v_{tm}
\end{bmatrix}
\overset{Transform\;to}{\rightarrow} 
\bar{S_i}=\begin{bmatrix}
a_1^{(i)} & a_2^{(i)} & ... & a_m^{(i)}
\end{bmatrix}
\label{eq:mvbm}
\end{split}
\end{equation}

Where $a_{k}^{(i)}=\frac{\sum_{j=1}^{t} v_{jk}^{(i)}}{t}$, for $k\in \{1,2,..m\}, i\in \{1,2,..n\}$.

In equation (\ref{eq:mvbm}), \(S_i\) represents the data of station $i$, and \(v_{jk}^{(i)}\) is the value of \(j^{th}\) month of \(k^{th}\) climate variable at station i. \(a_k^{(i)}\) is average of time series with length t of \(k^{th}\) climate variable for \(i^{th}\) station. The clustering algorithm suffers from different scales of the variables, as it is based on the distance between points. Therefore, data are normalized to maintain the performance of the algorithm. Finally, the spectral clustering algorithm with the specific number of clusters, is applied to the normalized data. 

\subsubsection{PCBM method}
The PCBM is similar to the MVBM, except that the data representation is different. In this approach, we have created a new feature space on the correlations between the seven time series at each station. This representation method of climate data is described by Steinhaeuser \textit{et al.} \citeyearpar{Steinhaeuser2009}. According to this method, for any two time series $TS_1$ and $TS_2$ of one data point, the Pearson correlation is computed. Therefore, for each data sample, we compute the correlation between all pairs of time series, 
\textit{i.e.} ${m \choose 2}$ pairs. In other words, each matrix is transformed into a vector with elements which range from $1$ to $-1$. Finally, the spectral clustering algorithm 
is applied to the new feature space. 

\section{Results and Discussion}
\label{sec:res}

The results are categorized into three parts. The first part is the output from applying the GPBM method (Algorithm \ref{algo:GPBM}) on the synoptic data of Fars province and its time complexity analysis. The second one is analyzing the generated clusters from GPBM and comparing the results with benchmarking methods and the known climate classifications. And the third part is the examination of seasonal climate classification produced by GPBM.

\subsection{Technical analysis of the proposed method}
To classify the climate of Fars, GPBM is implemented in Python on a PC with Intel Core i5 with $2.67$GHz processor and $4$GB memory and run on the synoptic data described in Section~\ref{subsec:data}. 

\begin{figure}[htb]
\centering
\includegraphics[width=.6\textwidth]{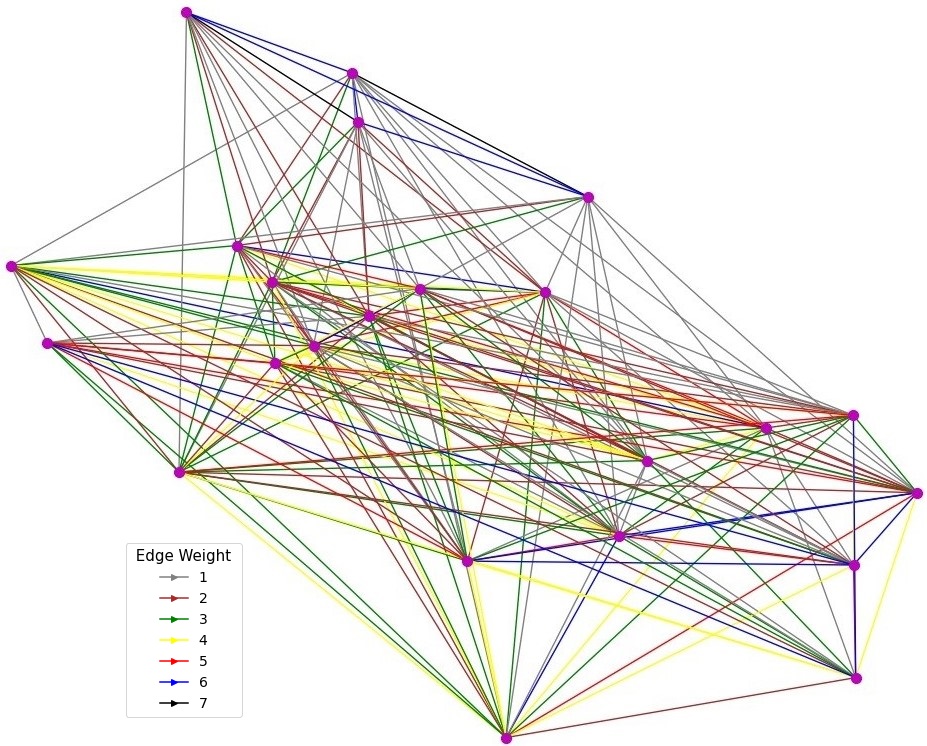}
\caption{The constructed graph based on defined similarity measure. The color of each edge indicates its weight, for example gray color refers to W=1 and brown color refers to W=2. (Best viewed in color.)}
\label{fig:graph}
\end{figure}

The algorithm parameters are set to $n=23$, $m=7$, $k=3$ and $\rho=0.7$. As mentioned in the previous section, we set the value of $k$ based on prior knowledge on the study area. According to the experts' opinion and also the number of samples of the problem which is $23$ stations, $k=3$ is a good choice. After executing Step 3 of GPBM, the initial graph $G$ is created. Figure~\ref{fig:graph} shows the constructed graph from the synoptic data of Fars. For legibility, the weight of the edges is determined with the different colors and each node is located in the graph according to its geographical location on the map. As shown in the figure, the graph is dense. In the graph $G$, the number of weak connections (\textit{i.e.} the edges with low weight) is considerable. Therefore, in the next step, by using the forward pruning technique with $\rho$=$0.7$, $\tau$ = $5$ is obtained. Consequently, edges with weight less than $5$ are removed.

\begin{figure}[htb!]
\centering
\includegraphics[width=.6\textwidth]{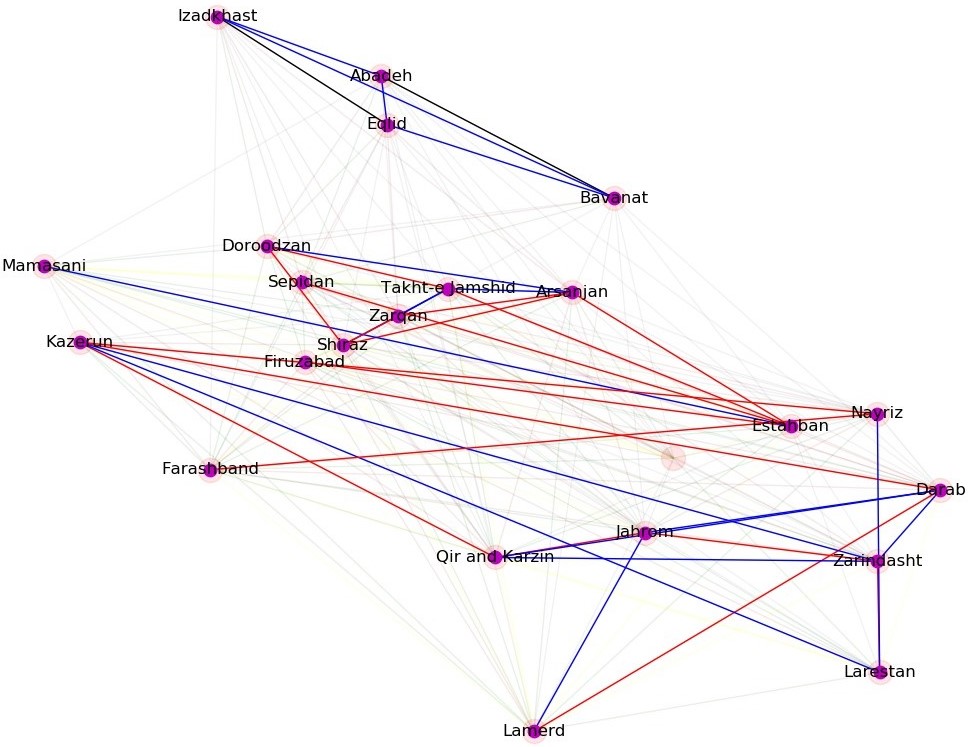}
\caption{The final graph after pruning with $\tau$=5}
\label{fig:prun_graph}
\end{figure}

The pruned graph $G'$ is illustrated in Fig. \ref{fig:prun_graph}. Finally, by partitioning the pruned graph, climate classification of region is determined. If we look closely at the $G'$ in Fig.~\ref{fig:prun_graph}, we will observe that this graph consists of three connected components that divide the area into three climate regions. In general, to partition the graph, we use the Fluid algorithm introduced by Pares \textit{et al.} \citeyearpar{Pares2018}, which takes the number of partitions as input, and we set it to $3$ as the value $k$ in k-means. 

The average execution time of the program on the mentioned database is equal to $2.34$ seconds. In order to determine the effect of different parameters of the algorithm on its execution time, a detailed complexity analysis is presented.

We can mention GPBM algorithm as consisting of five steps. The first step, data set production, is actually a pre-processing step that we leave out for complexity calculation. The second to fifth steps are time-consuming which must be executed sequentially. In the second step, the most time consuming operation is to call the clustering algorithm. The spectral clustering algorithm must be executed $m$ times, which $m$ is the number of climate variables. This part is performed in parallel. Therefore, the total cost of second step is equal to the time complexity of the clustering algorithm. John Scott \citeyearpar{Scott2000} expressed the time complexity of spectral clustering algorithm as $O(nzkt)$ where $z$ is the average number of rows in the similarity matrix and $t$ is the number of iterations required for k-means to converge. We can approximate $z$ with $n^2d$ where $d$ is length of time series. So, overall time complexity of second step is $O(n^3dkt)$. In graph construction step, an adjacency matrix with dimensions $n^2$ is constructed based on the similarities calculated in the previous step. The time complexity of this step is linear. To prune a graph, the forward pruning algorithm is executed. This method sorts the weighted edges and then selects the maximum $\rho n^2$ edges to create the graph. If we consider the initial graph as a complete graph in the worst case, (\textit{i.e.} the number of edges is $O(n^2)$), then the cost of this part is equal to $O(n^2 lg n + \rho n^2)$. And in the last step, the computational time complexity equals to the cost of graph partitioning algorithm. Since we use the Fluid algorithm, it is the number of edges of the pruning graph, $O(n^2)$.

In general, according to the above analysis, the execution time of the algorithm does not increase as the number of climate variables increases. 
In other words, the complexity of algorithm with more climate variables will remain constant. But, as the most computationally intensive step is the execution of spectral clustering, the number of stations is the main factor in the temporal complexity of the GPBM method. As the number of stations increases, the execution time of the algorithm will also increase. Furthermore, if the length of the time series becomes much longer, it will affect the execution time of the second stage of the algorithm. Of course, its effect is much less than the effect of the number of stations.

\subsection{Analyzing generated clusters}
\begin{figure}[htb]
\centering
    \subfloat[GPBM]{\includegraphics[width=0.33\textwidth]{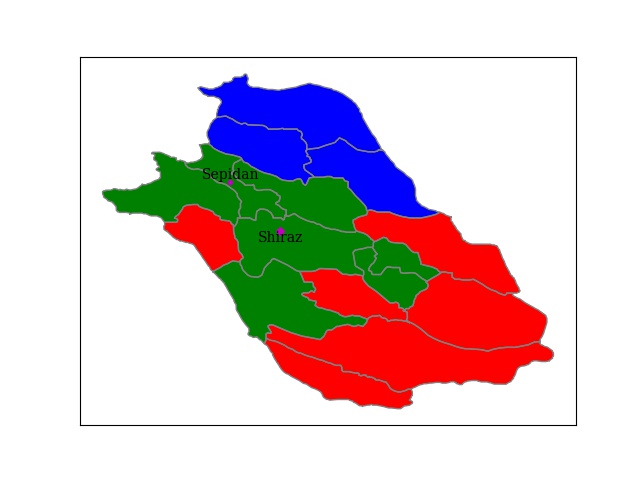}}
    \subfloat[MVBM]{\includegraphics[width=0.33\textwidth]{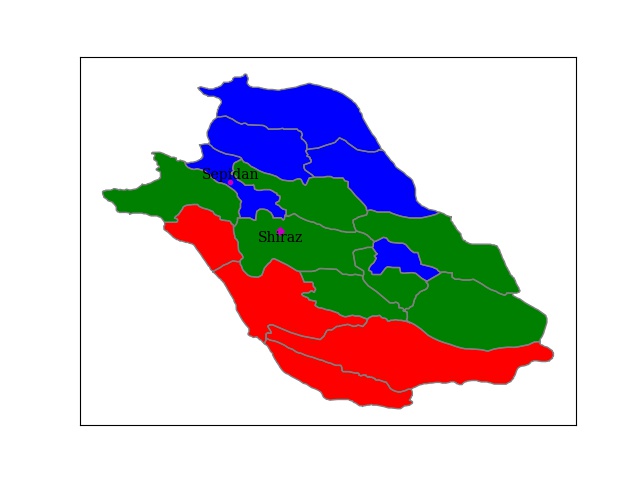}}
    \subfloat[PCBM]{\includegraphics[width=0.33\textwidth]{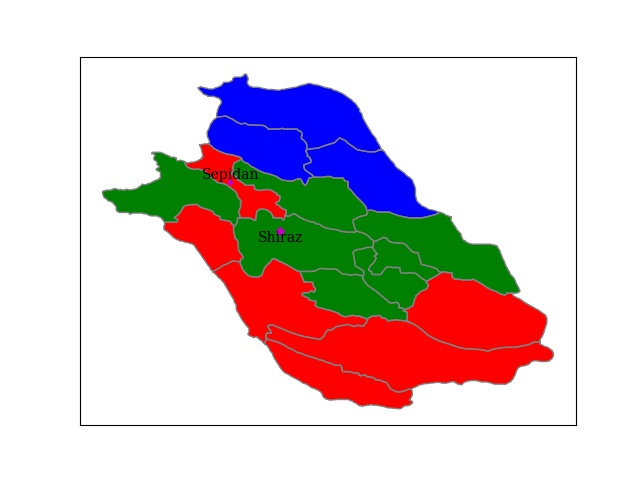}}
\caption{Climate classification result by proposed and benchmarking methods.(Blue: cluster1, Green: cluster2, Red: cluster3)}
\label{fig:maps}
\end{figure}

The classification of stations by the three described methods is written in Table \ref{tab:resultmethods} and illustrated in Fig.~\ref{fig:maps}. As it can be seen, the results obtained by GPBM are much more similar to the results of PCBM rather than that of MVBM. The reason for this, is the ability of both to preserve the spatiotemporal nature of the data. All three algorithms maintain the spatial aspect of the data by considering each station as one sample in the clustering algorithm. 
In terms of maintaining the temporal property of data, PCBM captures the temporal nature of the data to some extent as it calculates the Pearson coefficient between time series. 
The MVBM method, on the other hand, cannot preserve the temporal dimension of data with the mean value.

\begin{table*}[htb]
\begin{center}
\caption{The classification result of three proposed methods}
\label{tab:resultmethods}
\begin{tabular}{l P{3.5cm} P{4.5cm} P{4.5cm}}
\multicolumn{1}{l}{ \textit{Method} } &  \textit{Cluster 1}  &  \textit{Cluster 2}  &  \textit{Cluster 3}  \\ 
\hline
 \textit{MVBM}  & {\cellcolor[rgb]{0.8,0.8,0.8}}Abadeh, Eqlid, Izadkhast, Bavanat, Arsanjan, Estahban, Sepidan, Doroodzan & {\cellcolor[rgb]{0.8,0.8,0.8}}Jahrom, Darab, Zarqan, Nayriz, Shiraz, Mamasani, Takht-e Jamshid & {\cellcolor[rgb]{0.8,0.8,0.8}}Zarrin Dasht, Farashband, Kazerun, Firuzabad, Qir and Karzin, Lamerd, Larestan \\
 \textit{PCBM}  & Abadeh, Eqlid, Izadkhast, Bavanat & \textbf{Jahrom}, Zarqan, \textbf{Nayriz}, Shiraz, Mamasani, Takht-e Jamshid, Doroodzan, Estahban, Arsanjan & Zarrin Dasht, Farashband, Kazerun, \textbf{Firuzabad}, Qir and Karzin, Lamerd, Larestan, \textbf{Sepidan}, Darab \\
 \textit{GPBM}  & {\cellcolor[rgb]{0.8,0.8,0.8}}Abadeh, Eqlid, Izadkhast, Bavanat & {\cellcolor[rgb]{0.8,0.8,0.8}}Zarqan, Shiraz, Mamasani, Takht-e Jamshid, Doroodzan, Estahban, Arsanjan, \textbf{Sepidan, Firuzabad} & {\cellcolor[rgb]{0.8,0.8,0.8}}Zarrin Dasht, Farashband, Kazerun, Qir and Karzin, Lamerd, Larestan, \textbf{Nayriz, Jahrom}, Darab
\end{tabular}
\end{center}
\end{table*}

The GPBM and PCBM methods roughly classify the same stations in each cluster. The first cluster of them (\textit{i.e.} Cluster 1 in Table \ref{tab:resultmethods}) includes the northern stations of the province. In other words, they conclude the northern cities have a similar climate. The interesting note here is that this result is obtained without using the geographical coordinates. This demonstrate that the climate variables alone reflect the topographic and geographical features of the region they come from. However, there are significant differences in the other clusters. The differences are highlighted in Table \ref{tab:resultmethods}. The GPBM method places Sepidan and Firuzabad in Cluster 2 and Jahrom and Nayriz in Cluster 3, while the PCBM method performs the reverse.

One of the notable differences is found at the Sepidan station. It is marked on the map in Fig.~\ref{fig:maps}. We observe that each method has recognized a different climate for Sepidan. The MVBM places it in the Cluster 1 which includes Abadeh, Eqlid and Estahban. The PCBM method categorizes Sepidan, which has a cold weather (with the annual average temperature, \ang{16.68}C and the average precipitation, 42.43 mm in 2017), with tropical cities such as Larestan and Lamerd (with the annual average temperature, \ang{28.01}C and the average precipitation, 28.9 mm in 2017) as Cluster 3. And the GPBM method does not group Sepidan with northern cities and tropical cities. Based on this method, it can be concluded that the climate of Sepidan is similar to Shiraz (see the Cluster 2 in Table \ref{tab:resultmethods}). However, according to meteorological data and the history of these two cities, it is not true at least in winter. In the next section, we will investigate this issue on seasonal climate. 

We can conclude that due to the GPBM method incorporates all data values in the specified interval into its calculations, it is able to better observe the changes in climate variables over time and its results will be more reliable than the two others methods. In other words, the result of GPBM is more rational since month-to-month sequencing information is considered in the time series representation in GPBM while the vector representation in PCBM and MVBM does not consider it. 

Table \ref{tab:comparison} reports the classification of common stations in three methods including our proposed method (GPBM), the De Martonne approach and the method introduced by Hatami and Khoshhal \citeyearpar{Hatami2010}. According to the content published on Iran Meteorological Organization, the De Martonne classification divides the meteorological stations of Fars into three groups. A group includes Abadeh, Izadkhast, Jahrom, Larestan, Lamerd and Nayriz. The second group consists of Eqlid, Estahban, Bavant, Takht-e Jamshid, Zarqan, Doroodzan, Arsanjan, Shiraz and Kazerun, and finally, the Sepidan station is grouped as third cluster. As it is known, its classification is not consistent with the known climate of stations. For example, this method has identified the climate of Abadeh is identical with cities like Larestan and Lamerd. The reason is that in De Martonne classification, the annual temperature and precipitation are just used to calculate the moisture index and to identify the boundary of regions.

\begin{table}[htb]
\begin{center}
\caption{The comparison of GPBM with available methods based on common stations.}
\label{tab:comparison}
\begin{tabular}{l P{2.5cm} P{3.5cm} P{2.5cm}}
\multicolumn{1}{l}{ \textit{Method} } &  \textit{Cluster 1}  &  \textit{Cluster 2}  &  \textit{Cluster 3}  \\ 
\hline
 \textit{GPBM}  & {\cellcolor[rgb]{0.8,0.8,0.8}}Abadeh, Eqlid & {\cellcolor[rgb]{0.8,0.8,0.8}}Zarqan, Shiraz, Takht-e Jamshid, Doroodzan, Sepidan & {\cellcolor[rgb]{0.8,0.8,0.8}}Kazerun, Lamerd, Larestan, Jahrom, Darab \\
 \textit{De Martonne}  & Abadeh, Jahrom, Larestan, Lamerd & Eqlid, Zarqan, Shiraz, Takht-e Jamshid, Doroodzan, Kazerun, & Sepidan \\
 \textit{Hatami and Khoshhal}  & {\cellcolor[rgb]{0.8,0.8,0.8}}Abadeh, Eqlid & {\cellcolor[rgb]{0.8,0.8,0.8}}Zarqan, Shiraz, Takht-e Jamshid, Doroodzan, Sepidan & {\cellcolor[rgb]{0.8,0.8,0.8}}Kazerun, Jahrom, Lamerd, Larestan, Darab
\end{tabular}
\end{center}
\end{table}

Compared to the factor analysis method running on Fars, provided by Hatami and Khoshhal, the GPBM method shows similar results. They classified the province into four climatic areas. Sepidan, Doroodzan, Takht-e Jamshid, Zarqan and Shiraz build one group. Abadeh and Eqlid are the second climate group. Larestan, Lamerd and Darab are placed into third region group. The fourth region group consists of Kazerun and Jahrom. Since they considered the third and fourth regions as a tropical climate, regarding common stations in both types of research, 
their results are very close to the GPBM results. However, our proposed method does not require complex statistical computations of their method, and it can visualize the climate classification with graph representation.

\subsection{Analyzing seasonal climate classification}
\begin{figure}[htb]
  \begin{center}
    \subfloat[Spring]{\includegraphics[width=0.4\textwidth]{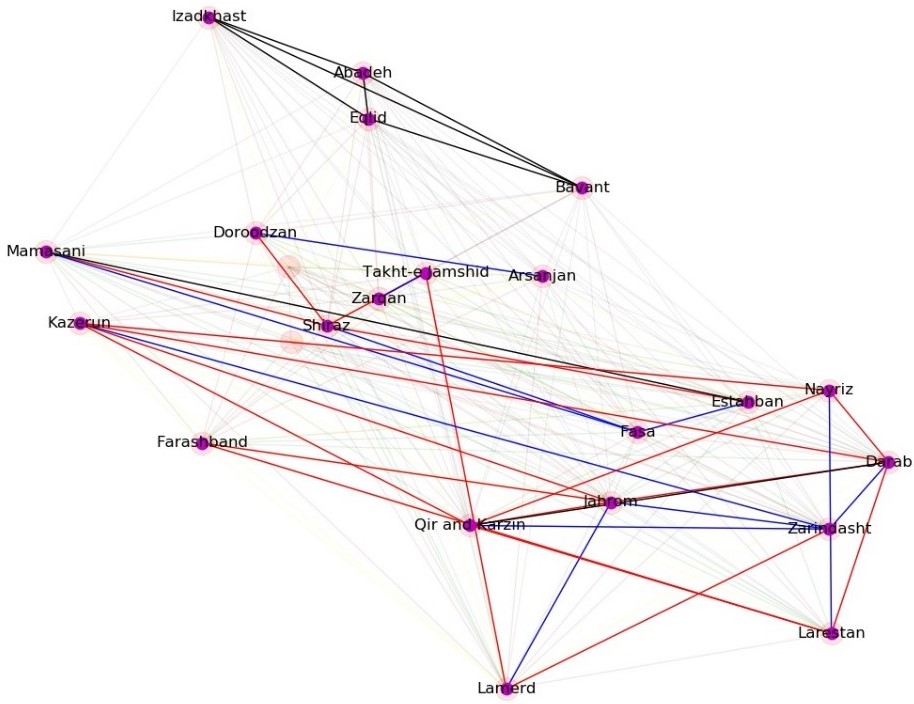}}
    \subfloat[Summer]{\includegraphics[width=0.4\textwidth]{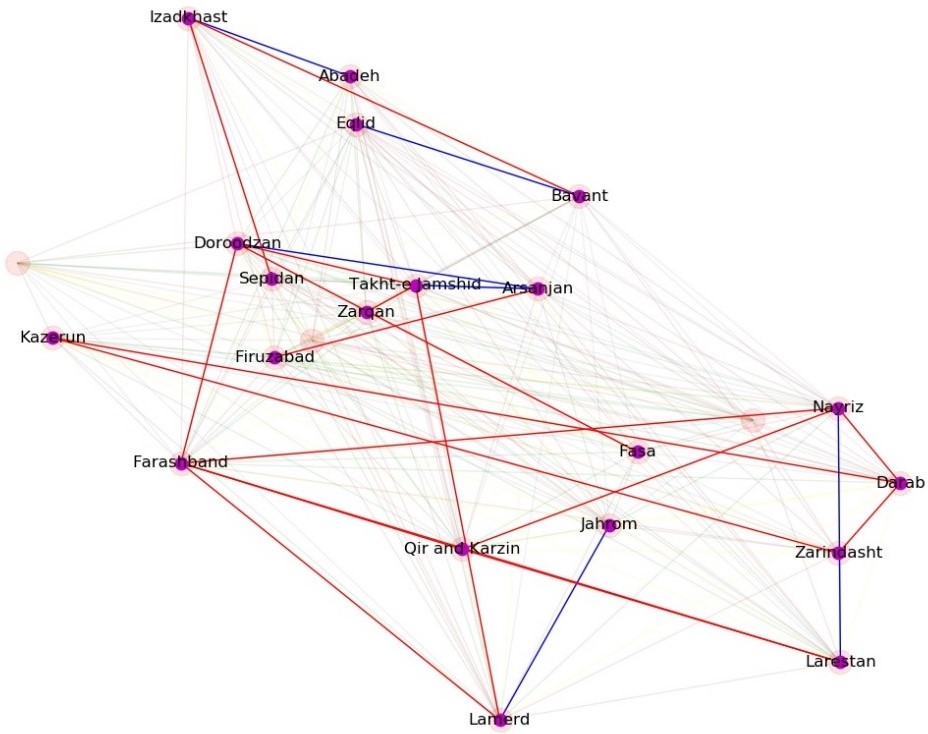}}\\
    \subfloat[Autumn]{\includegraphics[width=0.4\textwidth]{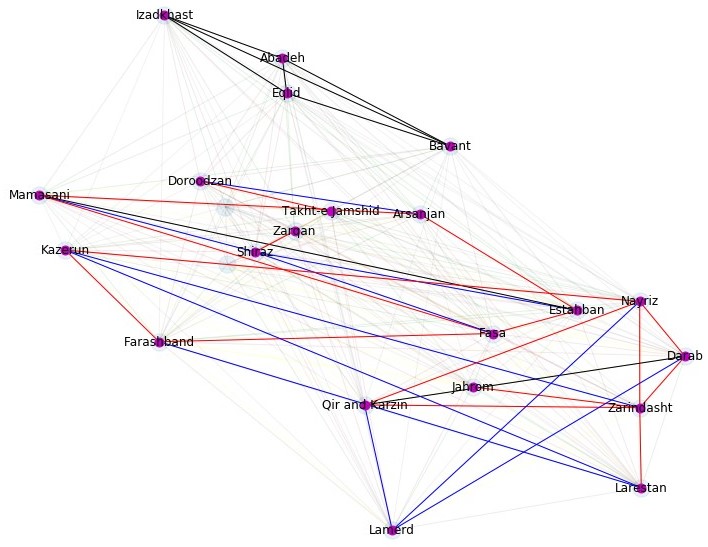}}
    \subfloat[Winter]{\includegraphics[width=0.4\textwidth]{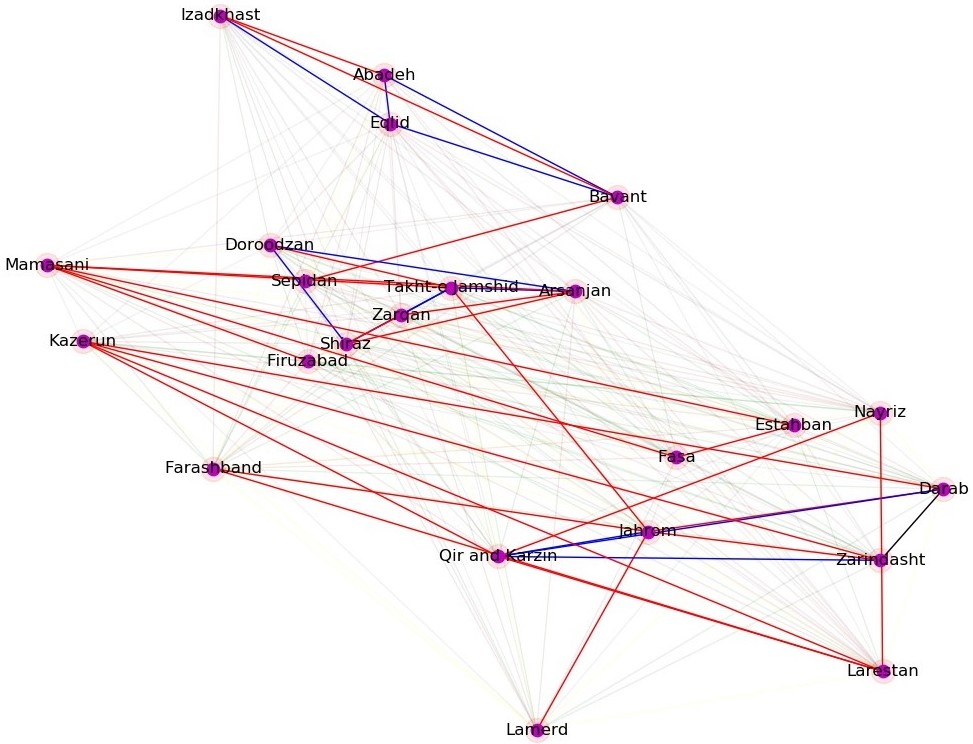}}
  \end{center}
  \caption{(a) Spring, (b) Summer, (c) Autumn, and (d) Winter climate classification of Fars.}
  \label{fig:climate-class}
\end{figure}

As we mentioned in the previous section, the climate of two regions, \textit{e.g.} Sepidan and Shiraz, may be similar in one season, yet they may differ in another season. Also, we know in the K\"oppen-Geiger method, the seasonal values of variables, such as summer and winter precipitation are used to determine the type of climate \citep{Raziei2017}. According to this point, we try to cover another aspect of climate variation using seasonal climate classification. To analyze this issue, we apply the GPBM method to seasonal data to perform the seasonal climate classification. To construct each seasonal graph, Algorithm~\ref{algo:GPBM} is applied to the corresponding seasonal data. The seasonal graphs are shown in Fig.~\ref{fig:climate-class}. 

First, we look at some of the characteristics of the seasonal graphs. Their statistical properties are listed in Table~\ref{tab:chrsGraph}. We note that $G'$ is the graph after the pruning stage. According to the numbers in this table, when the annual data is used, the number of edges in the initial graph is less than the number of edges obtained by the seasonal data. It can be concluded that the similarity of climate variables in different stations throughout the year are less. Furthermore, when climate classification is based on September, October, and November (\textit{i.e.} Autumn season), the graph has the maximum number of edges and maximum weight. This means that the climate conditions of the different regions of the province are more similar in Autumn, related to the variables used. 

\begin{table}[htb]
\centering
\caption{The characteristics of the created graphs.}
\label{tab:chrsGraph}
\arrayrulecolor[rgb]{0.788,0.788,0.788}
\begin{tabular}{l|P{1cm}|P{1cm}|P{1cm}|P{1cm}|P{1cm}|}
\textit{Season} & \#edges of G & Total weight of G & \#edges of $G'$ & Total weight of $G'$ & Density of $G'$ \\
\hline
 \textit{Spring}  & {\cellcolor[rgb]{0.929,0.929,0.929}}221 & {\cellcolor[rgb]{0.929,0.929,0.929}}622 & {\cellcolor[rgb]{0.929,0.929,0.929}}37 & {\cellcolor[rgb]{0.929,0.929,0.929}}212 & {\cellcolor[rgb]{0.929,0.929,0.929}}0.18 \\ 
\arrayrulecolor{black}\hline
 \textit{Summer}  & 224 & 587 & 24 & 126 & 0.13 \\ 
\hline
 \textit{Autumn}  & {\cellcolor[rgb]{0.929,0.929,0.929}}227 & {\cellcolor[rgb]{0.929,0.929,0.929}}651 & {\cellcolor[rgb]{0.929,0.929,0.929}}41 & {\cellcolor[rgb]{0.929,0.929,0.929}}218 & {\cellcolor[rgb]{0.929,0.929,0.929}}0.16 \\ 
\hline
 \textit{Winter}  & 226 & 612 & 34 & 196 & 0.16 \\ 
\hline
 \textit{Annual} & {\cellcolor[rgb]{0.929,0.929,0.929}}200 & {\cellcolor[rgb]{0.929,0.929,0.929}}583 & {\cellcolor[rgb]{0.929,0.929,0.929}}34 & {\cellcolor[rgb]{0.929,0.929,0.929}}189 & {\cellcolor[rgb]{0.929,0.929,0.929}}0.13 \\
\hhline{~>{\arrayrulecolor[rgb]{0.788,0.788,0.788}}|>{\arrayrulecolor{black}}->{\arrayrulecolor[rgb]{0.788,0.788,0.788}}|>{\arrayrulecolor{black}}->{\arrayrulecolor[rgb]{0.788,0.788,0.788}}|>{\arrayrulecolor{black}}->{\arrayrulecolor[rgb]{0.788,0.788,0.788}}|>{\arrayrulecolor{black}}->{\arrayrulecolor[rgb]{0.788,0.788,0.788}}|>{\arrayrulecolor{black}}->{\arrayrulecolor[rgb]{0.788,0.788,0.788}}|}
\hline
\end{tabular}
\end{table}

According to the graphs, the climate is more uniform in the summer rather than in the other seasons. One of the clusters clearly outlined in the summer graph includes Abadeh, Bavanat, Eqlid, Izadkhast and Sepidan. This cluster exists with more weight (indicating more similarity) in winter. Therefore, based on the summer and winter seasonal graph, the climate of Sepidan is the same as the Cluster 1 of GPBM in table \ref{tab:resultmethods}. Also, the northern cluster includes Abadeh, Bavanat, Eqlid and Izadkhast, without Sepidan. This cluster exists with maximum similarity (\textit{i.e.} the weight of every edges in the graph is $7$) in spring and autumn graph. As it can be seen from both graphs, Sepidan has been removed from the graphs, which means that the climate of Sepidan is not similar to other stations with respect to the threshold $\tau=5$. It shows that the result of De Martonne approach for Sepidan is not irrelevant. This is because the average data is close to the two spring and summer values, the output of the De Martonne is that the climate type of Sepidan is not like other cities.

For more investigation, the quantitative data of stations in this cluster, for spring and autumn is analyzed. Since the classical method (\textit{i.e.} De Martonne method in this work) takes into account the temperature and precipitation, we compare the value of these variables. Table \ref{tab:first_cluster} reports the average temperature \((T_{avg})\), maximum temperature \((T_{max})\), minimum temperature \((T_{min})\) and average of total precipitation \((Total Prec._{avg})\) for Cluster 1 and Sepidan station. The fluctuation of Sepidan temperature is in the range of stations for the first cluster except for minimum temperature. 
However, it is very different in terms of precipitation value. Therefore, based on these two variables, Sepidan cannot be placed in this cluster.

\begin{table*}[htb]
\centering
\caption{The synoptic data of first cluster and Sepidan station.}
\label{tab:first_cluster}
\arrayrulecolor{black}
\begin{tabular}{l!{\color{black}\vrule}l!{\color{black}\vrule}cccc!{\color{black}\vrule}!{\color{black}\vrule}cccc!{\color{black}\vrule}} 
\arrayrulecolor{black}\hline
\multicolumn{2}{c|}{\multirow{2}{*}{Station}} & \multicolumn{4}{c!{\color{black}\vrule}!{\color{black}\vrule}}{Spring (Mar., Apr. \& May)} & \multicolumn{4}{c|}{Autumn (Sep., Oct. \& Nov.)} \\ 
\cline{3-10}
\multicolumn{2}{c|}{} & \multicolumn{1}{l!{\color{black}\vrule}}{\(T_{avg}\)} & \multicolumn{1}{l!{\color{black}\vrule}}{\(T_{max}\)} & \multicolumn{1}{l|}{\(T_{min}\)} & \multicolumn{1}{l!{\color{black}\vrule}!{\color{black}\vrule}}{Total \(Prec._{avg}\)} & \multicolumn{1}{l!{\color{black}\vrule}}{\(T_{avg}\)} & \multicolumn{1}{l!{\color{black}\vrule}}{\(T_{max}\)} & \multicolumn{1}{l!{\color{black}\vrule}}{\(T_{min}\)} & \multicolumn{1}{l|}{Total \(Prec._{avg}\)} \\ 
\arrayrulecolor{black}\hline
\multirow{4}{*}{First Cluster} & Izadkhast & 14.83 & 32.2 & -9 & \multicolumn{1}{c||}{16.54} & 16.01 & 26.4 & 1.3 & 10.03 \\ 
\arrayrulecolor{black}\cline{2-2}
 & Abadeh & 14.63 & 35 & -8.8 & 10.78 & 14.98 & 27.42 & 1.56 & 8.13 \\ 
\cline{2-2}
 & \multicolumn{1}{l|}{Eqlid} & 13.00 & 32 & -10.6 & 34.63 & 13.91 & 25.2 & 1.11 & 14.69 \\ 
\cline{2-2}
 & \multicolumn{1}{l|}{Bavanat} & \multicolumn{1}{l}{15.56} & 32.4 & -7.8 & 28.2 & 16.58 & 25.96 & 1.35 & 11.32 \\ 
\hhline{>{\arrayrulecolor{black}}======>{\arrayrulecolor{black}}::>{\arrayrulecolor{black}}====>{\arrayrulecolor{black}}|}
\rowcolor[rgb]{0.906,0.902,0.902} \multicolumn{2}{c|}{Sepidan} & 14.69 & 32 & -7.8 & 54.05 & 17.26 & 25.6 & 6.37 & 35.84 \\
\hhline{>{\arrayrulecolor{black}}--->{\arrayrulecolor{black}}---|b|--->{\arrayrulecolor{black}}->{\arrayrulecolor{black}}|}
\end{tabular}
\arrayrulecolor{black}
\end{table*}

Hence, we can see that the findings in seasonal graphs are consistent with our observations. These graphs provide useful information about the climate of regions in different seasons. We anticipate that seasonal climate classification can be exploited in agricultural studies. This is because such studies 
involve the seasonal climate analysis 
to determine the appropriate areas for planting seasonal plants.

\section{Conclusion and Future Work}
\label{sec:conc}
In this paper, we proposed a new graph-based approach called GPBM, for local climate classification using synoptic data stored as time series. It performed the climate classification without changing the nature of climate data and removing information. The GPBM approach does not have the limitations of PCA and is acceptable in terms of computational cost. It does not depend on the limited number of variables, and it preserves the spatiotemporal nature of climate data. From the comparative analysis, GPBM presented the climate of stations of Fars more realistically. The results exhibited the effect of other climate variables 
on the geographical diversity of the region.

We also introduced the seasonal graphs for analyzing the seasonal variability of climate. Using GPBM, we can have a meaningful comparison of the climate of individual regions at different seasons. Based on the results, it is recommended to use GPBM as an analytical framework on any spatial scale with different climate variables for the study of the climate change. The proposed method is applied to the local regions of Iran, and the graph is constructed with synoptic data with 23 nodes. In the future work, we intend to assessing the scalability of the method for larger areas using grided data extracted from satellite images. We also intend to benchmark other clustering algorithms in the proposed framework. 
The graph-based climate classification algorithm needs to be further optimized for this purpose, because of the presence of the large number of nodes. We will also extend our work on visualizing larger domains spanning different continents in our future works.

\section*{Acknowledgment}
We would like to thank Dr.\ Mohammad Jafar Nazemosadat, head of the Atmospheric and Oceanic Research Center at Shiraz University, for his guidance in understanding the matter and analyzing the meteorological data. The ADAPT Centre  for  Digital Content Technology is funded under the SFI Research Centres Programme (Grant 13/RC/2106) and is co-funded under the European Regional Development Fund.

\section*{Disclaimer}
The author(s) declare(s) that there is no conflict of interest regarding the publication of this article. Also, none of the authors listed on the manuscript are employed by a government agency that has a primary function other than research and/or education.




\end{document}